\documentclass[reprint,prb,aps,twocolumn,superscriptaddress]{revtex4-1}

\usepackage[english]{babel}
\usepackage{graphicx}
\usepackage{color}

\newcommand{\MoSe}{{${\rm MoSe_2}$}}
\newcommand{\Moire}{Moir\'e}

\usepackage{enumitem,amssymb}
\newlist{todolist}{itemize}{2}
\setlist[todolist]{label=$\square$}
\usepackage{pifont}

\def \ETH{Institute for Quantum Electronics, ETH Z\"urich, CH-8093 Z\"urich, Switzerland}
\def \NIMS{National Institute for Materials Science, Tsukuba, Ibaraki 305-0044, Japan,\\ 
$^*$These authors contributed equally to this work.}

\begin{document}

\title{Moir\'e superlattice in a MoSe$_2$/hBN/MoSe$_2$ heterostructure: from coherent coupling of inter- and intra-layer excitons to correlated Mott-like states of electrons}

\author{Yuya Shimazaki$^*$}
\affiliation{\ETH}

\author{Ido Schwartz$^*$}
\affiliation{\ETH}

\author{Kenji Watanabe}
\affiliation{\NIMS}

\author{Takashi Taniguchi}
\affiliation{\NIMS}

\author{Martin Kroner}
\affiliation{\ETH}

\author{Ata\c{c} Imamo\u{g}lu}
\affiliation{\ETH}

\maketitle

{\bf Two dimensional materials and their heterostructures constitute a promising platform to study correlated electronic states as well as many body physics of excitons. Here, we present experiments that unite these hitherto separate efforts and show how excitons that are dynamically screened by itinerant electrons to form exciton-polarons, can be used as a spectroscopic tool to study interaction-induced incompressible states of electrons. The MoSe$_2$/hBN/MoSe$_2$ heterostructure that we study exhibits a long-period \Moire\ superlattice as evidenced by coherent-hole tunneling mediated avoided crossings between the intra-layer exciton with three inter-layer exciton resonances separated by $\sim 3$\,meV. For electron densities corresponding to half-filling of the lowest \Moire\ subband, we observe strong layer-paramagnetism demonstrated by an abrupt transfer of all $\sim 1500$ electrons from one MoSe$_2$ layer to the other upon application of a small perpendicular electric field. Remarkably, the electronic state at half-filling of each MoSe$_2$ layer is resilient towards charge redistribution by the applied electric field, demonstrating an incompressible Mott-like state of electrons. Our experiments demonstrate that optical spectroscopy provides a powerful tool for investigating strongly correlated electron physics in the bulk and pave the way for investigating Bose-Fermi mixtures of degenerate electrons and dipolar excitons.}

Van der Waals heterostructures incorporating transition metal dichalcogneide (TMD) bilayers open up new avenues for exploring strong correlations using  transport and optical spectroscopy. In contrast to similar structures in III-V semiconductors, these heterostructures exhibit possibilities for exotic material combinations, creation of \Moire\ superlattices exhibiting narrow electronic bands\cite{Yu2017, Wu2018a, Wu2018, Ruiz-Tijerina2019}, and strong binding of spatially separated inter-layer excitons\cite{Fang2014, Rivera2015, Jauregui2018, Calman2018, Ciarrocchi2019}. Recently, ground-breaking transport experiments in twisted bilayer graphene demonstrated a fascinating range of strongly correlated electron physics in a single system\cite{Cao2018, Cao2018a, Yankowitz2019, Liu2019, Sharpe2019, Lu2019, Serlin2019}: by varying the filling factor $\nu$ of the lowest energy \Moire\ from 0 to 1, the ground-state of the interacting electron or hole system could be reversibly changed from a superconductor for a large range of $\nu$ to a Mott insulator at $\nu=n/4$ ($n=1,2,3$) or a band insulator at $\nu=1$. In fact, this system realizes a two-dimensional (2D) Fermi-Hubbard model on a triangular lattice with a fully-tunable electron density -- a paradigmatic example of a strongly correlated electronic system with many open questions. 

In parallel, optical spectroscopy in van der Waals heterostructures have revealed the prevalence of many-body hybrid light-matter states, termed exciton-polarons\cite{Sidler2017, Efimkin2017}, in the excitation spectra of electron or hole doped monolayers.  Advances in material quality and device fabrication has lead to the observation of \Moire\ physics of non-interacting excitons in MoSe$_2$/WSe$_2$~\cite{Seyler2019, Tran2019}, MoSe$_2$/WS$_2$\cite{Alexeev2019}, and WS$_2$/WSe$_2$\cite{Jin2019} heterobilayers. Potential of this new system for investigating many-body physics was recently revealed in a remarkable demonstration of a long-lived inter-layer exciton condensate\cite{Fogler2014, Wang2019}. Here, we describe experiments in a heterostructure incorporating a MoSe$_2$/hBN/MoSe$_2$ homobilayer that in several ways combine the principal developments in these two fields to demonstrate interaction-induced incompressible states of electrons. We provide an unequivocal demonstration of hybridization of inter- and intra-layer excitons mediated by coherent hole tunneling\cite{Deilmann2018, Chaves2018, Ruiz-Tijerina2019, Alexeev2019, Gerber2019} between the two MoSe$_2$ layers: the avoided crossings that we observe in optical reflection not only show the formation of dipolar excitons with a strong optical coupling but also reveal the existence of at least 3 \Moire\ bands of indirect excitons. We then demonstrate that intra-layer exciton-polaron resonances provide a sensitive tool to investigate correlated electronic states in the bulk. Equipped with this spectroscopic tool, we observe strong layer-paramagnetism\cite{Zheng1997, Ezawa2000} and an incommpressible Mott-like state of electrons when each layer has half filling.

\section{Device structure and basic characterization}

\begin{figure*}[h!]
\centering
\includegraphics[width=1.0\textwidth]{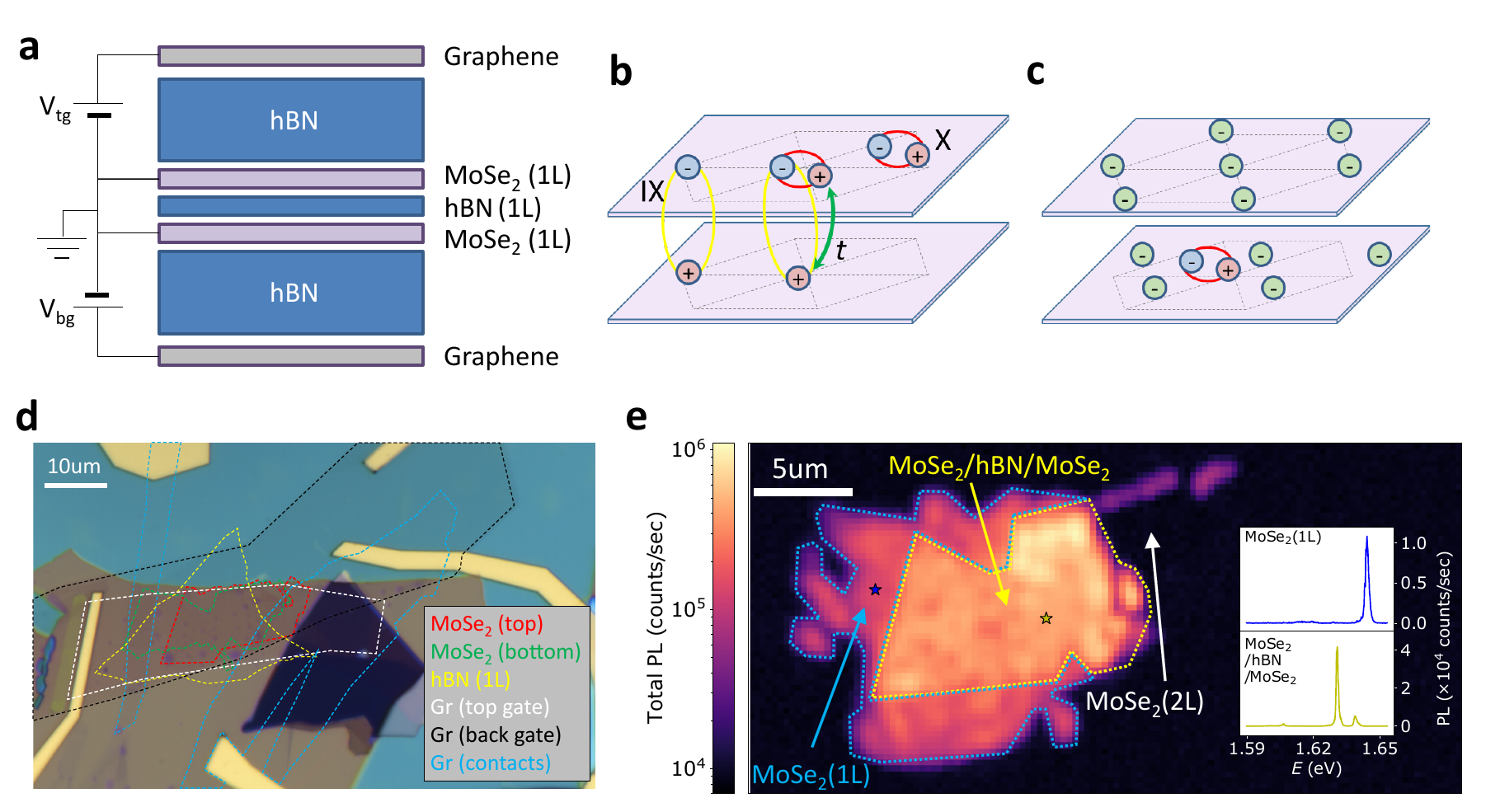}
\caption{{\bf Device structure and the basic characteristics.} 
{\bf a} Schematic image of the device structure. 
$V_{\rm tg}$ ($V_{\rm bg}$) is the applied voltage to the top (bottom) gate.
{\bf b, c} Schematic image of coupled inter- and intra-layer exciton ({\bf b})
and electrons in the \Moire\ lattice probed by the intra-layer exciton ({\bf c}).
Purple planes correspond to \MoSe\ layers, and dashed lines indicate a \Moire\ unit cell.
The pink (light blue) circles with $+$ ($-$) sign indicate holes (electrons) forming excitons, and the electron-hole pair enclosed by the red (yellow) ellipse indicates intra-layer (inter-layer) exciton.
The green double arrow in {\bf b} indicates tunnel coupling of holes
through the monolayer hBN barrier.
The light green circles in {\bf c} indicate electrons filling the \Moire\ lattice.
{\bf d} Optical microscope image of the device.
The border of each flake is highlighted with dashed lines,
and the material is indicated in the gray box
with the corresponding color. (The abbreviation "Gr" stands for Graphene.)
{\bf e} Spatial map of the integrated photoluminescence from 1.59eV to 1.65eV.
The blue and yellow dashed lines indicate the boundary of
the area of monolayer \MoSe\ and \MoSe/hBN/\MoSe, respectively.
The inset shows representative PL spectra of monolayer \MoSe\ and \MoSe/hBN/\MoSe\ measured at
the positions indicated with the blue and the yellow stars in the main figure,
respectively.
} \label{fig1}
\end{figure*}

We show the schematic of the device structure in Fig. \ref{fig1}a.
By using a double gate structure, we can control the electric field and the chemical potential of the device independently.
Few-layer graphene serves as transparent gates,
top and bottom hBN serve as gate insulators, and
middle hBN (monolayer) serves as a tunnel barrier.
The crystal axis of the \MoSe\ layers are aligned to be close to 0 degree
using the tear-and-stack technique\cite{Kim2016}.
Both \MoSe\ layers are grounded via few-layer graphene contacts.
Fig. \ref{fig1}d shows the optical microscope image of the device.
Fig. \ref{fig1}b is a schematic image of a dipolar exciton formed by coherent coupling of
inter-layer exciton (IX) and intra-layer exciton (X) via hole tunneling.
Fig. \ref{fig1}c shows a schematic image of the electrons in a \Moire\ lattice
probed by intra-layer exciton in low electron density regime.

Figure~\ref{fig1}e shows a spatial map of total photoluminescence (PL) from the device.
Here, both top and bottom gate voltages are kept at zero Volts.

We observe PL from regions with monolayer \MoSe, but not from bilayer \MoSe,
where two \MoSe\ flakes are in direct contact
(the area around the point indicated by the white arrow in Fig. \ref{fig1}e).
On the other hand, the \MoSe/hBN/\MoSe\ area shows bright PL.
This indicates that the heterostructure becomes a direct band gap system 
owing to the reduction of the inter-layer hybridization of the valence bands at the $\Gamma$ point\cite{Zhang2014}, due to the presence of monolayer hBN.
Typical PL spectra of the monolayer \MoSe\ and the \MoSe/hBN/\MoSe\ area
are shown in the inset of Fig. \ref{fig1}e: there are pronounced intra-layer exciton luminescence peaks in both areas. Even though the heterostructure is fabricated from a single \MoSe\ layer using tear-and-stack technique, we observe two distinct intra-layer exciton peaks in the \MoSe/hBN/\MoSe\ region. We observe that this strain-induced energy difference~\cite{He2013, Conley2013} between the PL from the top and bottom layers varies across the the sample (see Supplementary Information S2).

\section{Coherent inter-layer hole tunneling and dipolar excitons}

\begin{figure*}[h!] %[!t]
\centering
\includegraphics[width=1.0\textwidth]{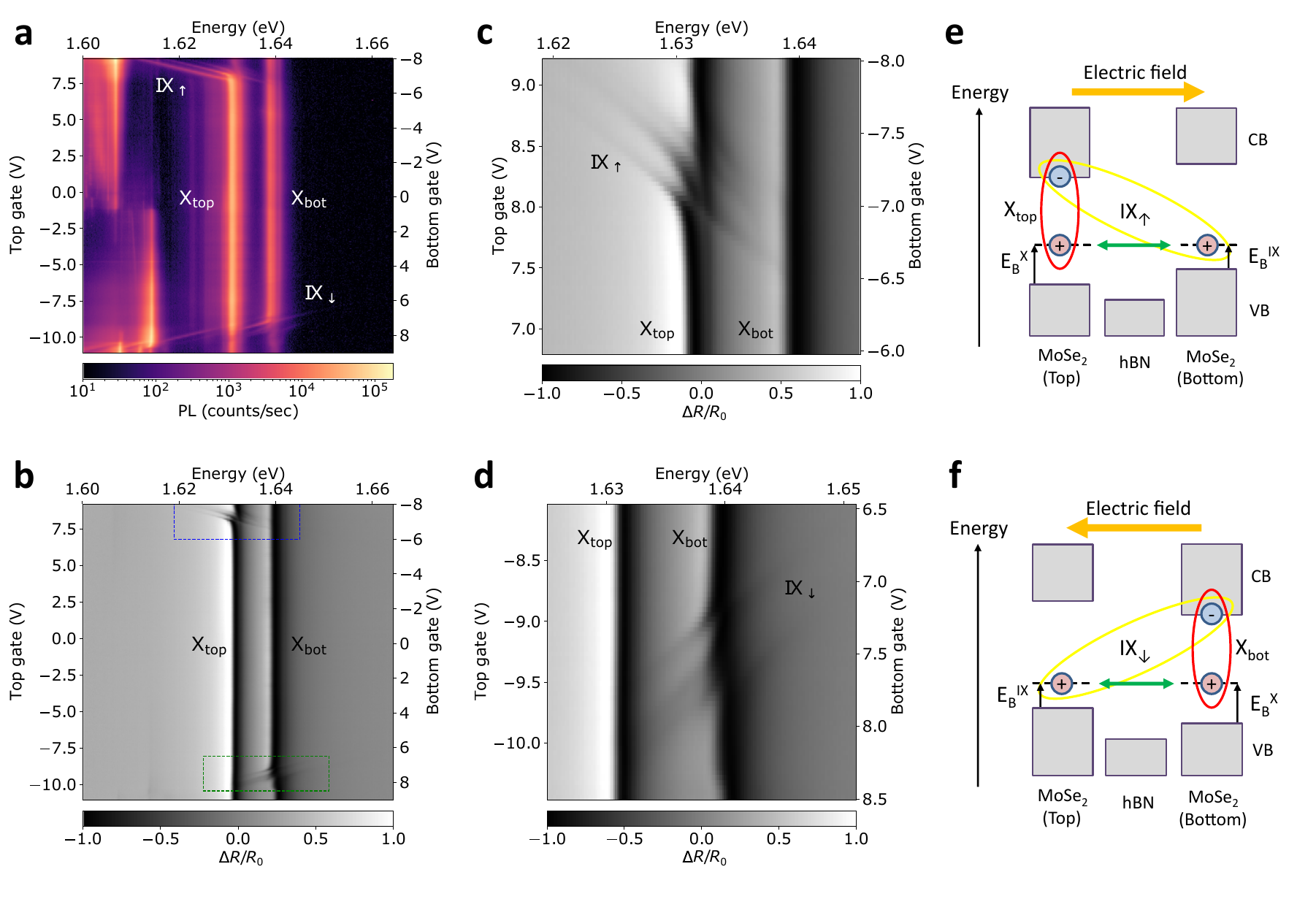}
\caption{{\bf Electric field dependence of photoluminescence and differential reflectance at charge neutrality.}
Gate dependence of photoluminescence {\bf a} and differential reflectance {\bf b} of $\rm MoSe_2/hBN/MoSe_2$.
Top and bottom gate voltages are scanned together to tune the electric field
at a constant chemical potential 
(scanned along the dashed line L4 shown in Fig. \ref{fig3}c).
The intensity of the photoluminescnece {\bf a} is shown in log scale.
{\bf c}, {\bf d} Magnified plots of {\bf b}.
The corresponding area of {\bf c} and {\bf d} is indicated by the blue and green dashed rectangulars
in {\bf b}, respectively.
{\bf e}, {\bf f} Schematic of the energy bands and the exciton energy alignment
under electric fields at the hole resonant conditions of excitons.
} \label{fig2}
\end{figure*}

We first analyze the electric field ($E_z$) dependence of the elementary optical excitations of the \MoSe/hBN/\MoSe\ section in the absence of itinerant electrons or holes. To this end, we scan the top and bottom gate voltages (along L4 indicated in Fig. \ref{fig3}c) together to change $E_z$
while keeping the homo-bilayer system in the charge neutral regime. The PL spectrum we thus obtain is depicted in Fig. \ref{fig2}a: using the top ($V_{\rm tg}$) and bottom ($V_{\rm bg}$) gate voltage dependence, we determine that the PL spectra around 1.632\,eV and 1.640\,eV stem from intra-layer exciton in top and bottom layer ($\rm X_{top}$ and $\rm X_{bot}$), respectively. For high values of $|E_z|$ depicted in the top and bottom parts of the color-coded PL spectrum, we observe PL lines with a strong $E_z$ dependence: we identify these PL lines as originating from inter-layer excitons with a large dipole moment leading to a sizeable Stark shift.

The spectra for positive (negative) $V_{\rm tg}$ regime corresponds to the inter-layer exciton $\rm IX_{\uparrow}$ ($\rm IX_{\downarrow}$) which has a hole in the bottom (top) layer and an electron in the top (bottom) layer. The associated dipole-moment of the inter-layer exciton changes its polarity for $V_{\rm tg}$~$\sim 0$. By extrapolating the $\rm IX_{\uparrow}$  and $\rm IX_{\downarrow}$ PL lines and finding their crossing point, we estimate the energy difference between the inter- and the intra-layer exciton resonances at $E_z = 0$, which allows us to determine their binding energy difference to be $\sim$ 100\,meV.

Figure~\ref{fig2}b shows the differential reflectance ($\Delta R/R_0$) spectrum obtained for the same range of gate voltage scan as that of Fig. \ref{fig2}a. Here, $\Delta R / R_0 \equiv (R - R_0)/R_0$, with $R$ and $R_0$ denoting the reflectance signal from the \MoSe/hBN/\MoSe\ region, and background reflectance, respectively.
In accordance with the PL data (Fig. \ref{fig2}a), we find $\rm X_{top}$ and $\rm X_{bot}$ resonances around 1.632\,eV and 1.640\,eV, respectively. Moreover, for $V_{\rm tg}$ $\gtrsim$ 7.5\,V ($V_{\rm tg}$ $\lesssim$ -7.5\,V), we observe $\rm IX_{\uparrow}$ ($\rm IX_{\downarrow}$) hybridizing exclusively with $\rm X_{top}$ ($\rm X_{bot}$). Figures~\ref{fig2}c and \ref{fig2}d show the magnified plots of the regions highlighted with blue and green dashed lines in Fig.~\ref{fig2}b, confirming avoided crossing of an intra-layer exciton line with multiple inter-layer excitons. We first note that the observation of a sizeable reflection signal from $\rm IX_{\uparrow}$ away from the avoided crossing suggests that it is possible to resonantly excite long-lived inter-layer excitons in these structures. The hybridization of $\rm IX_{\uparrow}$ lines with $\rm X_{top}$, together with the lack of an avoided crossing with $\rm X_{bot}$ in Fig.~\ref{fig2}c, unequivocally shows that avoided crossings originate exclusively from coherent hole tunneling schematically shown in Fig. \ref{fig2}e.
Our observation, proving that the hole tunnel coupling is much larger than that of the electron, is consistent with the band alignment expected from first principle band-structure calculations\cite{Ozcelik2016}. This conclusion is also confirmed by the data depicted in Fig. \ref{fig2}d, where avoided crossing originates from coherent-hole-tunneling induced hybridization of $\rm IX_{\downarrow}$ and $\rm X_{bot}$
schematically shown in Fig. \ref{fig2}f.

One of the most remarkable features of the spectra depicted in Figs.~\ref{fig2}c~and~\ref{fig2}d is the existence of multiple avoided crossings associated with three inter-layer exciton resonances separated in energy by $\sim 3$\,meV. This inter-layer exciton fine-structure demonstrates the existence of a \Moire\ superlattice~\cite{Yu2017, Wu2018, Ruiz-Tijerina2019, Alexeev2019, Seyler2019, Tran2019, Jin2019}, originating from a small twist angle between the two \MoSe\ layers. The presence of an hBN tunnel barrier strongly suppresses the strength of the associated \Moire\ potential, rendering it sizeable only for the inter-layer excitons~\cite{Yu2017}.

\section{Charge configuration detection by exciton-polaron spectroscopy}

It is well established that presence of itinerant charges drastically alters the optical excitation spectrum~\cite{Xu2014}. Recent theoretical and experimental work established that the modified spectrum originates from dynamical screening of excitons by electrons or holes~\cite{Sidler2017, Efimkin2017}, that lead to the formation of a lower energy attractive polaron (AP) branch. Concurrently, the exciton resonance evolves into a repulsive polaron (RP) (see Supplementary Information S1). The particularly strong sensitivity of the RP resonance energy to changes in electron density renders it an ideal spectroscopic tool for monitoring, or sensing,
the electron density $n$ in the same layer~\cite{Back2017, Smolenski2019}.
The strain induced resonance energy difference between $\rm X_{top}$ and $\rm X_{bot}$, ensuring different energies for the corresponding $\rm RP_{top}$ and $\rm RP_{bot}$, together with the much weaker sensitivity of $\rm RP_{top}$ ($\rm RP_{bot}$) on $n$ in the bottom (top) layer, allows us to determine the charging configuration of the two layers simultaneously.
Since we are predominantly interested in the low carrier density regime
where the quasi-particle (bare-exciton) weight of the RP resonance is
close to unity, we will refer to it as the exciton resonance.

\begin{figure*}[h!] %[!t]
\centering
\includegraphics[width=1.0\textwidth]{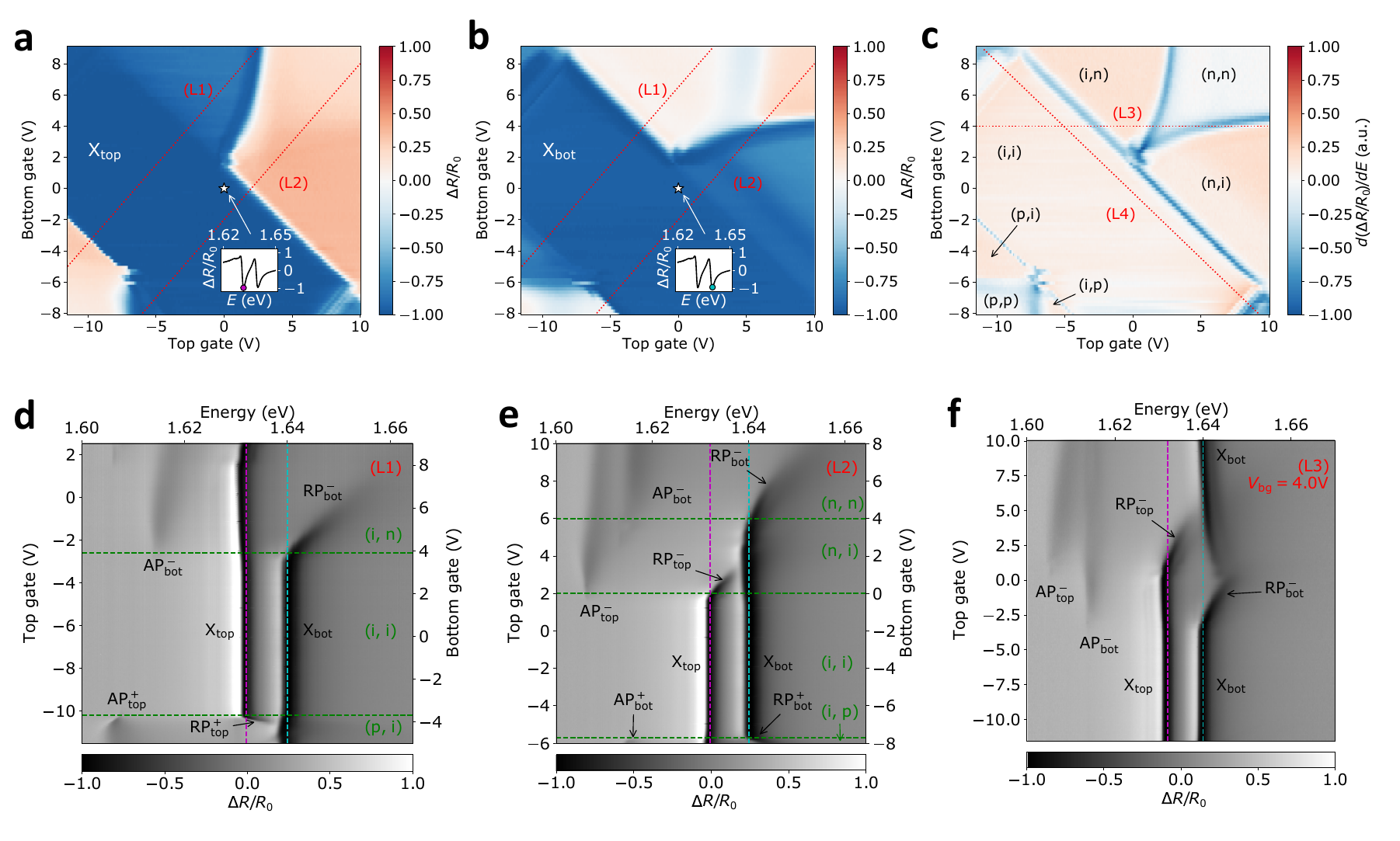}
\caption{{\bf Gate depedndence of differential reflectance spectrum of $\bf MoSe_2/hBN/MoSe_2$.}
{\bf a, b} Two gates dependence maps of differential reflectance
around top ({\bf a}) and bottom ({\bf b}) intra-layer exciton resonances
($E = 1.632 \rm eV$ and $E = 1.640 \rm eV$, respectively).
The insets of {\bf a} and {\bf b} show the differential reflectance spectrum
at $(V_{\rm tg}, V_{\rm bg}) = (0{\rm V}, 0{\rm V})$
(indicated with the white stars in the maps).
The magenta and cyan dots in the insets indicate the points
where $E = 1.632 \rm eV$ and $E = 1.640 \rm eV$, respectively.
{\bf c} Charge configuration diagram obtained by derivative of the differential reflectance spectrum
with respect to energy (sum of the derivatives at $E = 1.632 \rm eV$ and $E = 1.640 \rm eV$).
The charge configuration for each layer is indicated by p, i, n
which correspond to hole doped, neutral, electron doped, respectively,
and shown in the order of (top, bottom).
{\bf d} - {\bf f} Gate dependence of differential reflectance along the dashed lines L1 ({\bf d}), L2 ({\bf e}), and L3 ({\bf f})
shown in {\bf a}, {\bf b} and {\bf c}.
Magenta and cyan dashed lines indicate the top ($E = 1.632 \rm eV$) and bottom ($E = 1.640 \rm eV$) 
exciton resonance energies, respectively.
$\rm AP_{L}^{C}$ and $\rm RP_{L}^{C}$
stand for intra-layer attractive and repulsive polarons,
where $\rm L = "top"\ or\ "bot"$ stands for top or bottom layer,
and $\rm C = +\ or\ -$ stands for hole or electron as Fermi sea carriers.
Charge configuration is written in green
together with the green dashed lines indicating the charge configuration transition
point.
} \label{fig3}
\end{figure*}

Figures~\ref{fig3}a and \ref{fig3}b show the gate voltage dependence of $\Delta R/R_0$
at $E = 1.632\,{\rm eV}$ and $E = 1.640\,{\rm eV}$, which correspond to
the top ($\rm X_{top}$) and bottom ($\rm X_{bot}$) intra-layer exciton resonance energy
in the charge neutral regime, respectively. The inset to these figures show a line cut through the dispersive neutral exciton reflection spectrum, indicating the exciton energies at which we monitor $\Delta R/R_0$. 
Since a small increase of $n$ from $\sim 0$ to $1 \times 10^{11}$\,cm$^{-2}$ results in a change of $\Delta R/R_0$ from $\sim -1$ to $\gtrsim 0$, the blue areas in Figs.~\ref{fig3}a and \ref{fig3}b  correspond to the charge neutral regime of each layer. The red and white areas in turn, correspond to the electron or hole doped regime of each layer. This all-optical determination of the electrically resolved charge map of the bilayer provides an invaluable tool for monitoring the bulk properties of 2D materials which is not directly accessible in transport measurements.

To enhance the sensitivity of the charge map to the transition between the undoped and doped regimes and to visualize the charge configuration of both layers at the same time, we first evaluate the derivative of $\Delta R/R_0$ with respect to energy at $E = 1.632\,\rm eV$ and $E = 1.640\,\rm eV$, and then overlay $d (\Delta R/R_0)/d E$ obtained for both layers. The resulting charge map, depicted in Fig.~\ref{fig3}c, is closely reminiscent of the charging plateaus used to characterize gate-defined quantum dots~\cite{Hanson2007}. Since $d (\Delta R/R_0)/d E$ is only sensitive to changes in $n$, the blue regions in Fig.~\ref{fig3}c correspond to the regime where the charge configuration changes, allowing us to clearly separate the regions (t,b) where the top or bottom layer is neutral (t=i or b=i), electron doped (t=n or b=n) or hole doped (t=p or b=p).

We show typical gate voltage dependence of $\Delta R/R_0$ in Fig. \ref{fig3}d and
\ref{fig3}e which are obtained when the two gate voltages are scanned in a coordinated manner along the lines L1 and L2, indicated in Fig. \ref{fig3}a and \ref{fig3}b, respectively. In both plots, we confirm the emergence of the AP resonance and the associated blue shift of the exciton / RP energy around the charge configuration transition points, confirming the assignment obtained from $d (\Delta R/R_0)/d E$ in~Fig.~\ref{fig3}c. 

In stark contrast to the case of monolayer \MoSe\ (Supplementary Information S1), we find that the top (bottom) gate  dependence obtained by fixing the bottom (top) gate voltage, is not monotonic because of the screening of the applied gate voltage when one of the two layers is already doped. 
The observed responsivity to the applied gate voltages is consistent with the interpretation that the lower (higher) energy exciton resonance at $E = 1.632\,\rm eV$ ($E = 1.640\,\rm eV$) is $\rm X_{top}$ ($\rm X_{bot}$).
For example in Fig. \ref{fig2}b, by fixing $V_{\rm tg}$ = 0\,V and sweeping $V_{\rm bg}$ from negative to positive, we find drastic change of $\Delta R/R_0$ around $V_{\rm bg} \simeq$ 2\,V.
On the other hand, by fixing $V_{\rm bg}$ = 0\,V and sweeping $V_{\rm tg}$ from negative to positive, we find much less change of $\Delta R/R_0$.
From this asymmetry of the gate dependence, we can confirm that 
the resonance in Fig. \ref{fig2}b is originating from $\rm X_{bot}$.

Figure~\ref{fig3}f shows the gate voltage scan along L3, indicated in Fig. \ref{fig3}c, where we fixed $V_{\rm bg}$ at $4\,\rm{V}$ and scanned $V_{\rm tg}$. By sweeping $V_{\rm tg}$ from negative to positive, we find that the bottom layer gets electron doped around $V_{\rm tg} = -3\,\rm{V}$, and then gets depleted by increasing $V_{\rm tg}$ further.
This observation shows that electrons are transferred from the bottom layer to the top layer whilst  electrons are introduced into the top layer and  $V_{\rm bg}$ is kept unchanged. This counter-intuitive dependence shows up as the curving of the lines separating the charge configurations (n,i) and (n,n) in Fig. \ref{fig3}c. Similar inter-layer charge transfer behavior was previously observed in transport experiments in bilayer semiconductor systems \cite{Eisenstein1994, Larentis2014, Fallahazad2016} and was attributed to the negative compressibility. To the best of our knowledge, our experiments provide the first observation of negative compressibility, arising from dominance of intra-layer exchange interactions over kinetic energy, using optical spectroscopy.

\section{Interaction induced incompressible states}

The results we present in Sec.~II establish the existence of a \Moire\ superlattice for inter-layer excitons. On the one hand, the underlying periodic modulation of the electronic bands should lead to \Moire\ subbands for electrons (holes) in the conduction (valence) band. On the other hand, the absence of coherent electron tunneling indicates that the resulting subbands in the top and bottom layers do not hybridize. Taking into account the relatively strong conduction-band spin-orbit coupling in MoSe$_2$, the homobilayer system we are  investigating realizes a rather unique system exhibiting flat bands with layer and valley-spin degree of freedom; while the degeneracy associated with the former can be tuned using a perpendicular electric field ($E_z$), the latter can be controlled using a magnetic field ($B_z$). Moreover, our observation of negative compressibility (Sec.~III) indicates that the electron-electron interaction energy scale dominates over kinetic energy even at relatively high electron densities ($n \simeq 1 \times 10^{12}$\,cm$^{-2}$) where several \Moire\ bands in one layer are occupied. In this section, we explore electron correlation effects in the more interesting regime of low carrier densities by zooming in to the low-$n$ section of the charging map (Fig.~3c) where the (i,i), (i,n), (n,i) and (n,n) regions coalesce. The high sensitivity of the exciton/RP resonance energy, as well as the AP oscillator strength, to changes in electron density once again forms the backbone of our investigation. 

\begin{figure*}[h!] %[!t]
\centering
\includegraphics[width=1.0\textwidth]{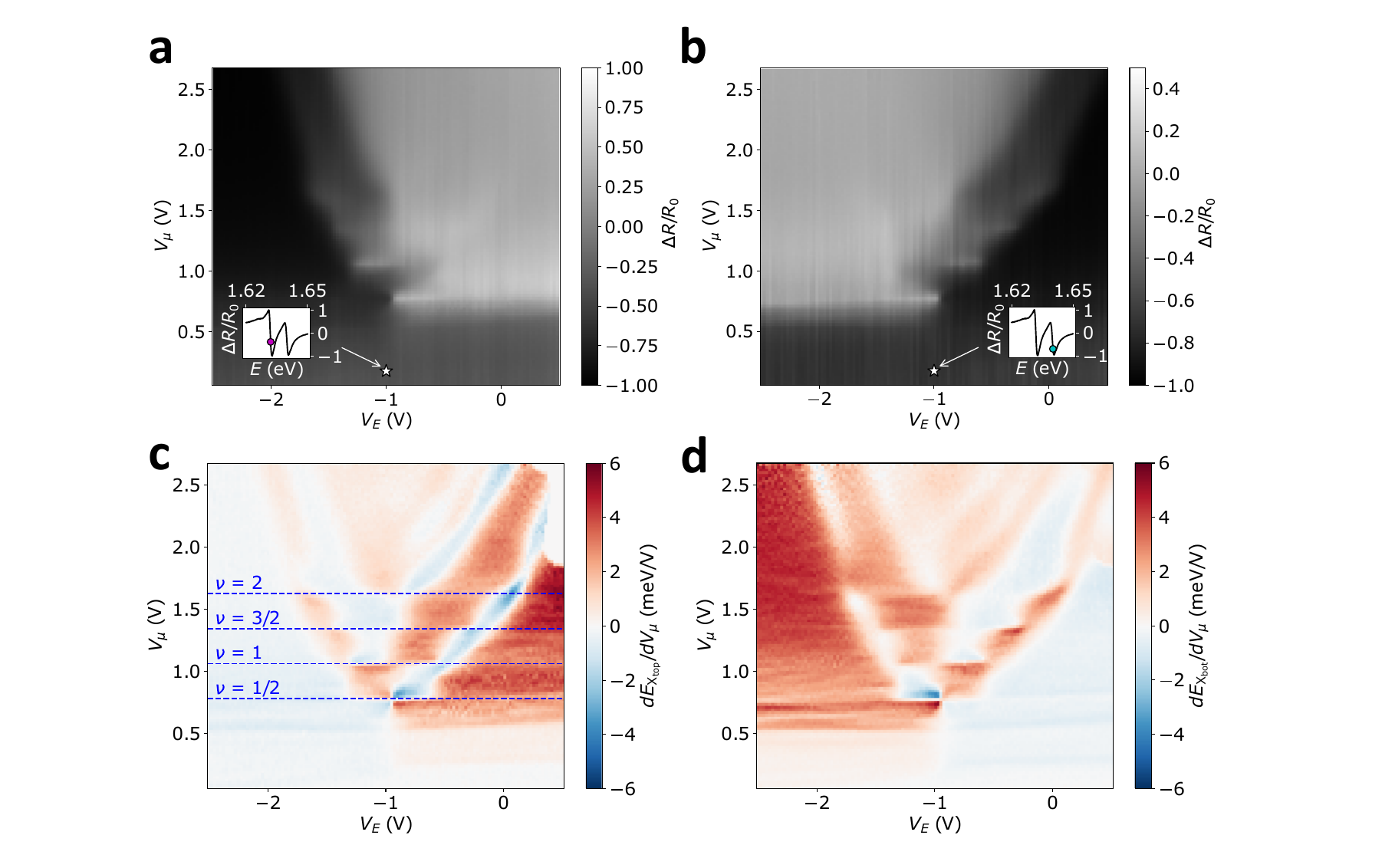}
\caption{{\bf Gate dependence of intra-layer exciton resonances in low electron density regime.}
{\bf a, b} Gate dependence maps of differential reflectance
around top ({\bf a}) and bottom ({\bf b}) intra-layer exciton resonances
($E = 1.6320 \rm eV$ and $E = 1.6402 \rm eV$, respectively).  The insets of {\bf a} and {\bf b} show the differential reflectance spectrum
at $(V_{E}, V_{\mu}) = (-1{\rm V}, 0.175{\rm V})$ (indicated with the white stars in the maps).
{\bf c, d} Gate dependence maps of top ({\bf c}) and bottom ({\bf d}) intra-layer exciton resonance energy differentiated by $V_{\mu}$.
} \label{fig4}
\end{figure*}

Figures \ref{fig4}a and \ref{fig4}b show the gate voltage dependence of differential reflectance close to the $\rm X_{top}$ and $\rm X_{bot}$ resonance energy for $n=0$ at $E_{\rm X_{top}}^0 = 1.6320$\,eV and $E_{\rm X_{bot}}^0 = 1.6402$\,eV, respectively. In these maps, a shift of the exciton resonance energy due to a change in $n$ is detected as a modification
of the exciton reflectance. The specific choice of $E_{\rm X_{top}}^0$ and $E_{\rm X_{bot}}^0$, indicated by the magenta and cyan points in the insets of Fig. \ref{fig4}a and \ref{fig4}b, maximizes the sensitivity to $n$. Instead of showing the reflectance map as a function of the top and bottom gate voltages, we now choose the horizontal and vertical voltage axes to be given by $V_E = 0.5 V_{\rm tg} - 0.5 V_{\rm bg}$ and $V_{\mu} = (7/15) V_{\rm tg} + (8/15) V_{\rm bg}$.
With this choice, vertical ($V_{\mu}$ axis) and horizontal ($V_{E}$ axis) cuts through the reflectance map leave $E_z$ and $\mu$ unchanged respectively, where $\mu$ denotes chemical potential.

Figures~\ref{fig4}a and \ref{fig4}b show a periodic modulation of the RP differential reflectance as a function of $V_{\mu}$, particularly in the low $n$ regime. Moreover, the modulation of the top and bottom layer reflectance are correlated and symmetric with respect to the $V_E = -1$\,V axis, indicating that for this value of $V_E$, the energy detuning of
the top and bottom $\rm MoSe_2$ layers is zero.
To gain further insight into the structure of correlated changes, we first determine the excitonic resonance energy for the top and bottom layers ($E_{\rm X_{top}}$ and $E_{\rm X_{bot}}$) by fitting the reflectance spectrum with two dispersive Lorentzian lineshapes (see Supplementary Information S3) and then plot the derivative of $E_{\rm X_{top}}$ and $E_{\rm X_{bot}}$ with respect to $V_{\mu}$ in 
Figs. \ref{fig4}c and \ref{fig4}d. The resulting map shows a remarkable checkerboard pattern that is complementary for the top and bottom layers. Since the blue shift of $\rm X_{top}$ and $\rm X_{bot}$ resonance while increasing $V_{\mu}$
(positive $dE_{\rm X_{top}}/dV_{\mu}$ and $dE_{\rm X_{bot}}/dV_{\mu}$)
corresponds to filling of electrons in the top and bottom layer respectively,
the complementary checkerboard like pattern indicates a layer by layer filling of electrons.
Note that a similar diagram has been reported in layer resolved capacitance measurement of Landau lavels in bilayer graphene\cite{Hunt2017}.

The observed periodicity in Fig.~\ref{fig4} evidences the existence of Moir\'e subbands for electrons.
In anticipation of the subsequent discussion,
we define a layer filling factor $\nu_{\rm L}$
(L = "top" or "bot" indicating top or bottom layer)
so that $\nu_{\rm L} = 1/2$ corresponds to 1 electron per Moir\'e
unit cell of a single layer, and a total filling factor $\nu$ as
$\nu = \nu_{\rm top} + \nu_{\rm bot}$.
From a capacitive model of our device,
we determine that $\nu = 1/2$ coincides with a remarkably
low electron density of $n = 2 \times 10^{11} {\rm cm^{-2}}$.
At this low electron density, $r_s$ parameter, which describes the ratio of interaction energy to kinetic energy, is estimated to be $r_s \simeq 14$.
The density periodicity corresponds to a \Moire\ superlattice latice constant of $\lambda_{\rm Moir\acute{e}} = 24\, {\rm nm}$ by assuming a triangular superlattice.
We indicate the values of $\nu$ corresponding to
$\nu$ = 1/2, 1, 3/2, 2 with blue dashed lines in Fig. \ref{fig4}c.

\begin{figure*}[h!]
\centering
\includegraphics[width=1.0\textwidth]{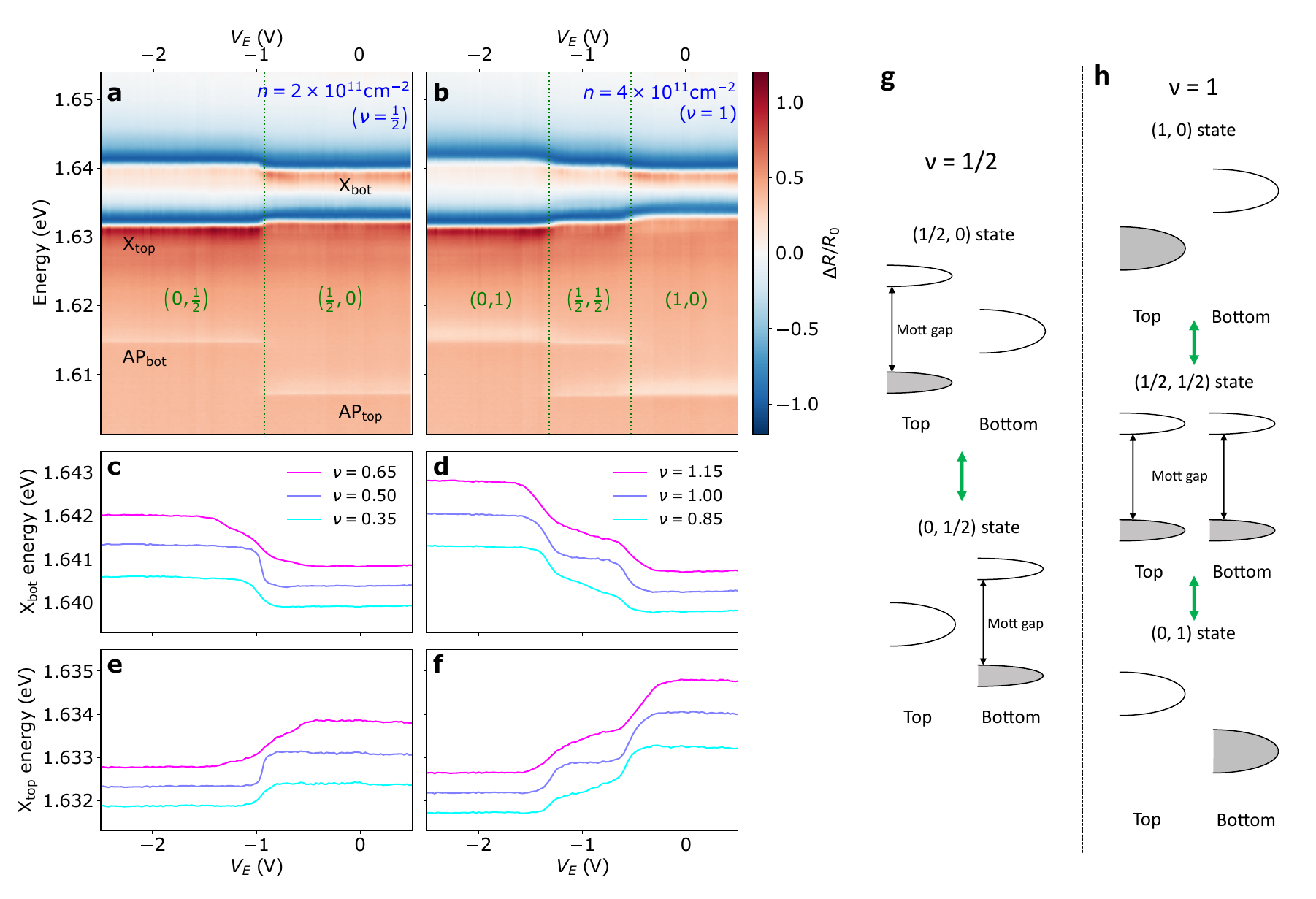}
\caption{{\bf Electric field dependence of differential reflectance spectrum in low electron density regime.}
{\bf a, b} Electric field ($V_E$) dependence of differential reflectance spectrum
for each fixed $V_{\mu}$ at $\nu = 1/2$ and $\nu = 1$.
Charging configuration of top and bottom layer is indicated by 
$(\nu_{\rm top}, \nu_{\rm bot})$ in green.
{\bf c, d}  $V_E$ dependence of $\rm X_{bot}$ resonance energy around each total filling of $\nu = 1/2$ ({\bf c}), $\nu = 1$ ({\bf d}).
{\bf e, f}  $V_E$ dependence of $\rm X_{top}$ resonance energy around each total filling of $\nu = 1/2$ ({\bf e}), $\nu = 1$ ({\bf f}). In {\bf c} - {\bf f}, the cyan curves are without offset and other curves are displaced by 0.5meV and 1.0meV.
{\bf g, h} Schematic picture of charge configuration with density of states
at filling of $\nu$ = 1/2 ({\bf g}) and $\nu = 1$ ({\bf h}).
} \label{fig5}
\end{figure*}

Figures~\ref{fig5}a and \ref{fig5}b show the $V_E$ dependence of the differential reflectance spectrum
for fixed $V_{\mu}$ where $\nu$ = 1/2 ($n = 2 \times 10^{11} {\rm\,cm^{-2}}$)
and $\nu$ = 1 ($n = 4 \times 10^{11}\,{\rm cm^{-2}}$), respectively. In Fig. \ref{fig5}a, we find an abrupt shift of exciton energy together with complete oscillator strength transfer between $\rm AP_{top}$ and $\rm AP_{bot}$. This shows that the electrons are completely and abruptly transferred from one layer to the other layer upon changing $E_z$. Figs. \ref{fig4}c and \ref{fig4}e show the extracted $\rm X_{bot}$ and $\rm X_{top}$ energies
around $\nu$ = 1/2. Remarkably, the abrupt jump in excitonic resonance is pronounced at $\nu$ = 1/2,
and smeared out for both lower ($\nu$ = 0.35) and higher filling factors ($\nu$ = 0.65).
These measurements show that abrupt transfer of practically all of the $\sim 1500$ electrons within the region we monitor optically is linked to the emergence of an interaction-induced incompressible state
in the lowest \Moire\ subband at $\nu$ = 1/2 filling. As the filling factor is increased (decreased) towards $\nu =1/2$, the electronic system shows an ever stronger layer-paramagnetism, due to enhanced role of interactions, but otherwise exhibits a continuous inter-layer transfer of electrons as a function of $E_z$ that would be expected from a compressible state. Close to $\nu=1/2$, there is a phase transition to an incompressible state that can be accommodated either in the top or bottom layer (see Fig. \ref{fig5}g).

Figure~\ref{fig5}b shows that for $\nu =1$, the polaron reflectance spectrum as a function of $E_z$ is characterized by 3 plateau-like regions. We attribute the abrupt jumps in the excitonic resonance energy to the transition from $(\nu_{\rm top}, \nu_{\rm bot})$ = (0, 1), through (1/2, 1/2), to (1, 0) configurations (see Fig. \ref{fig5}h).
This explanation is confirmed by the corresponding changes in the oscillator strength of the AP resonances of the top and bottom layers. In (1, 0) and (0, 1) configurations, we measure a reflectance signal either from $\rm AP_{top}$ or $\rm AP_{bot}$, consistent with full layer polarization. In the (1/2, 1/2) configuration, we find the oscillator strength of $\rm AP_{top}$ and $\rm AP_{bot}$ to be identical and equal to half the value obtained under (1,0) for $\rm AP_{top}$. 
The extracted excitonic resonance energy $\rm X_{bot}$ and $\rm X_{top}$ around $\nu = 1$
is shown in Figs.~\ref{fig5}d and \ref{fig5}f, respectively.
The plateau structure of the (1/2, 1/2) state with abrupt jumps in $n$ for $V_E = -1.2$\,V and $V_E = -0.7$\,V is clearly visible at $\nu = 1$, but is smeared out for both lower ($\nu = 0.85$) and higher fillings ($\nu = 1.15$).
We conclude that as the total $n$ is changed away from $\nu = 1$, both top and bottom layers become
compressible, showing smooth changes in layers occupancy as a function of small $E_z$ values ($-1.1\,{\rm V} < V_E < 0.8$\,V).
The emergence of the stabilized (1/2, 1/2) plateau at $\nu = 1$ strongly suggests that there is mutual stabilization of the incompressible electronic state due to the inter-layer electron-electron interactions. If the $\nu_{\rm top} = 1/2$ or  $\nu_{\rm bot} = 1/2$ had been incompressible for $\nu = 0.85$, we would have observed corresponding plateaus in the the excitonic resonance reflection map. The reflectance data for higher fillings ($\nu = 3/2$ and $\nu = 2$)
are shown in the supplementary information (Fig. S5): in stark contrast to the (1/2, 1/2) configuration at $\nu = 1$, a plateau at the (1, 1) electron configuration is missing at $\nu = 2$ filling, indicating that the state with the corresponding integer fillings is not sufficiently stabilized by the inter-layer interactions.

Finally, we emphasize that our identification of $n = 2 \times 10^{11}\,{\rm cm^{-2}}$ yielding half-filling of a single-layer \Moire\ subband is supported by our measurements at $B_z = 7$\,T: In Fig.~S6 we observe that the plateau structure observed for $\nu=2$ under full valley polarization of electrons is identical to that observed for $ B_z=0$\,T, even though the total number of electronic states per \Moire\ subband is halved due to giant valley-spin susceptibility of electrons in MoSe$_2$\cite{Back2017}. This observation shows that the incompressibility is determined by filling of each \Moire\ site by a single electron, irrespective of its degeneracy.

\section{Discussion}
The experiments we describe in Sec.~IV demonstrate the existence of Mott-like incompressible electronic states for half-filling of the lowest \Moire\ subband. Unlike prior reports\cite{Tran2019, Alexeev2019, Seyler2019, Jin2019}, our experiments are carried out for long \Moire\ superlattice latice constant of $\lambda_{\rm Moir\acute{e}} = 24\, {\rm nm}$ and $r_s$ parameter of $r_s  \simeq 14$. The weakness of the \Moire\ potential stemming from the hBN layer separating the two MoSe$_2$ layers in turn ensures that the on-site interaction strength is comparable to, or possibly larger than, the depth of the \Moire\ potential. In this sense, the homobilayer system realizes a rather unique regime of the Fermi-Hubbard model where some of the expectations such as an antiferromagnetic ground-state need not be applicable. 

In addition to establishing twisted TMD homo-bilayers as a promising system for investigating Mott-Wigner physics\cite{Imada1998, Camjayi2008, Zarenia2017, Wu2018} originating from strong electronic correlations, our experiments open up new avenues for exploring interactions between dipolar excitons and electrons confined to flat bands. In particular, the structure we analyzed could be used to realize and study Bose-Fermi mixtures consisting of degenerate electrons strongly interacting with an exciton condensate generated by resonant laser excitation. The phase diagram of such a mixutre is currently not fully understood~\cite{Ludwig2011} but is expected to provide a rich playground for many-body physics, including but not limited to exciton-mediated superconductivity~\cite{Little1964, Ginzburg1970, Laussy2010, Cotlet2016}. 

{\it Note added} After completion of this work, we became aware of two
manuscripts, arXiv:1910.08673 and arXiv:1910.09047, reporting similar results in a different material system.

\newpage

%% Figure1

\newpage

%% Figure2

\newpage

%% Figure3

\newpage

%% Figure4

\newpage

%% Figure5

\newpage

\section*{METHODS}
All \MoSe, graphene, and hBN flakes are obtained by mechanical exfoliation of bulk crystals.
The flakes are assembled together using the dry transfer technique\cite{Wang2013} in an Ar filled glove box.
The crystal axis of top and bottom \MoSe\ layers are aligned to be close to 0 degree
using tear-and-stack technique\cite{Kim2016}.
The metal electrodes to graphene layers are formed by Ti/Au (5\,nm/145\,nm).
The contact to the bottom graphene gate is formed by Cr/Au (3\,nm/147\,nm) 
using the one-dimensional contact technique\cite{Wang2013}
by etching the hBN layer with reactive ion etching using $\rm CHF_3/O_2$ as mixture gas.

The photoluminescence measurements were performed using a HeNe laser (633nm) as an excitation source.
The reflectance measurements were performed using a single mode fiber coupled broadband LED
with a center wavelength of 760nm and a bandwidth of 20nm.
In both photoluminescence and reflectance measurements, we used a long working distance apochromatic
objective lens with NA = 0.65 (attocube LT-APO/LWD/VISIR/0.65).
All optical spectroscopy measurements have been performed at cryogenic temperature ($T \sim 4 {\rm K}$). 

%% References
\bibliographystyle{naturemag.bst}
\bibliography{References}

\begin{thebibliography}{10}
\expandafter\ifx\csname url\endcsname\relax
  \def\url#1{\texttt{#1}}\fi
\expandafter\ifx\csname urlprefix\endcsname\relax\def\urlprefix{URL }\fi
\providecommand{\bibinfo}[2]{#2}
\providecommand{\eprint}[2][]{\url{#2}}

\bibitem{Yu2017}
\bibinfo{author}{Yu, H.}, \bibinfo{author}{Liu, G.-B.}, \bibinfo{author}{Tang,
  J.}, \bibinfo{author}{Xu, X.} \& \bibinfo{author}{Yao, W.}
\newblock \bibinfo{title}{{Moir{\'{e}} excitons: From programmable quantum
  emitter arrays to spin-orbit–coupled artificial lattices}}.
\newblock \emph{\bibinfo{journal}{Sci. Adv.}} \textbf{\bibinfo{volume}{3}},
  \bibinfo{pages}{e1701696} (\bibinfo{year}{2017}).

\bibitem{Wu2018a}
\bibinfo{author}{Wu, F.}, \bibinfo{author}{Lovorn, T.} \&
  \bibinfo{author}{MacDonald, A.~H.}
\newblock \bibinfo{title}{{Theory of optical absorption by interlayer excitons
  in transition metal dichalcogenide heterobilayers}}.
\newblock \emph{\bibinfo{journal}{Phys. Rev. B}} \textbf{\bibinfo{volume}{97}},
  \bibinfo{pages}{035306} (\bibinfo{year}{2018}).

\bibitem{Wu2018}
\bibinfo{author}{Wu, F.}, \bibinfo{author}{Lovorn, T.}, \bibinfo{author}{Tutuc,
  E.} \& \bibinfo{author}{MacDonald, A.~H.}
\newblock \bibinfo{title}{{Hubbard Model Physics in Transition Metal
  Dichalcogenide Moir{\'{e}} Bands}}.
\newblock \emph{\bibinfo{journal}{Phys. Rev. Lett.}}
  \textbf{\bibinfo{volume}{121}}, \bibinfo{pages}{026402}
  (\bibinfo{year}{2018}).

\bibitem{Ruiz-Tijerina2019}
\bibinfo{author}{Ruiz-Tijerina, D.~A.} \& \bibinfo{author}{Fal'ko, V.~I.}
\newblock \bibinfo{title}{{Interlayer hybridization and moir{\'{e}}
  superlattice minibands for electrons and excitons in heterobilayers of
  transition-metal dichalcogenides}}.
\newblock \emph{\bibinfo{journal}{Phys. Rev. B}} \textbf{\bibinfo{volume}{99}},
  \bibinfo{pages}{125424} (\bibinfo{year}{2019}).

\bibitem{Fang2014}
\bibinfo{author}{Fang, H.} \emph{et~al.}
\newblock \bibinfo{title}{{Strong interlayer coupling in van der Waals
  heterostructures built from single-layer chalcogenides}}.
\newblock \emph{\bibinfo{journal}{Proc. Natl. Acad. Sci.}}
  \textbf{\bibinfo{volume}{111}}, \bibinfo{pages}{6198--6202}
  (\bibinfo{year}{2014}).

\bibitem{Rivera2015}
\bibinfo{author}{Rivera, P.} \emph{et~al.}
\newblock \bibinfo{title}{{Observation of long-lived interlayer excitons in
  monolayer MoSe2-WSe2 heterostructures}}.
\newblock \emph{\bibinfo{journal}{Nat. Commun.}} \textbf{\bibinfo{volume}{6}},
  \bibinfo{pages}{6242} (\bibinfo{year}{2015}).

\bibitem{Jauregui2018}
\bibinfo{author}{Jauregui, L.~A.} \emph{et~al.}
\newblock \bibinfo{title}{{Electrical control of interlayer exciton dynamics in
  atomically thin heterostructures}}  (\bibinfo{year}{2018}).
\newblock \eprint{arXiv:1812.08691}.

\bibitem{Calman2018}
\bibinfo{author}{Calman, E.~V.} \emph{et~al.}
\newblock \bibinfo{title}{{Indirect excitons in van der Waals heterostructures
  at room temperature}}.
\newblock \emph{\bibinfo{journal}{Nat. Commun.}} \textbf{\bibinfo{volume}{9}},
  \bibinfo{pages}{1895} (\bibinfo{year}{2018}).

\bibitem{Ciarrocchi2019}
\bibinfo{author}{Ciarrocchi, A.} \emph{et~al.}
\newblock \bibinfo{title}{{Polarization switching and electrical control of
  interlayer excitons in two-dimensional van der Waals heterostructures}}.
\newblock \emph{\bibinfo{journal}{Nat. Photonics}}
  \textbf{\bibinfo{volume}{13}}, \bibinfo{pages}{131--136}
  (\bibinfo{year}{2019}).

\bibitem{Cao2018}
\bibinfo{author}{Cao, Y.} \emph{et~al.}
\newblock \bibinfo{title}{{Correlated insulator behaviour at half-filling in
  magic-angle graphene superlattices}}.
\newblock \emph{\bibinfo{journal}{Nature}} \textbf{\bibinfo{volume}{556}},
  \bibinfo{pages}{80--84} (\bibinfo{year}{2018}).

\bibitem{Cao2018a}
\bibinfo{author}{Cao, Y.} \emph{et~al.}
\newblock \bibinfo{title}{{Unconventional superconductivity in magic-angle
  graphene superlattices}}.
\newblock \emph{\bibinfo{journal}{Nature}} \textbf{\bibinfo{volume}{556}},
  \bibinfo{pages}{43--50} (\bibinfo{year}{2018}).

\bibitem{Yankowitz2019}
\bibinfo{author}{Yankowitz, M.} \emph{et~al.}
\newblock \bibinfo{title}{{Tuning superconductivity in twisted bilayer
  graphene}}.
\newblock \emph{\bibinfo{journal}{Science}} \textbf{\bibinfo{volume}{363}},
  \bibinfo{pages}{1059--1064} (\bibinfo{year}{2019}).

\bibitem{Liu2019}
\bibinfo{author}{Liu, X.} \emph{et~al.}
\newblock \bibinfo{title}{{Spin-polarized Correlated Insulator and
  Superconductor in Twisted Double Bilayer Graphene}}  (\bibinfo{year}{2019}).
\newblock \eprint{arXiv:1903.08130}.

\bibitem{Sharpe2019}
\bibinfo{author}{Sharpe, A.~L.} \emph{et~al.}
\newblock \bibinfo{title}{{Emergent ferromagnetism near three-quarters filling
  in twisted bilayer graphene}}.
\newblock \emph{\bibinfo{journal}{Science}} \textbf{\bibinfo{volume}{365}},
  \bibinfo{pages}{605--608} (\bibinfo{year}{2019}).

\bibitem{Lu2019}
\bibinfo{author}{Lu, X.} \emph{et~al.}
\newblock \bibinfo{title}{{Superconductors, Orbital Magnets, and Correlated
  States in Magic Angle Bilayer Graphene}}  (\bibinfo{year}{2019}).
\newblock \eprint{arXiv:1903.06513}.

\bibitem{Serlin2019}
\bibinfo{author}{Serlin, M.} \emph{et~al.}
\newblock \bibinfo{title}{{Intrinsic quantized anomalous Hall effect in a
  moir{\'{e}} heterostructure}}  (\bibinfo{year}{2019}).
\newblock \eprint{arXiv:1907.00261}.

\bibitem{Sidler2017}
\bibinfo{author}{Sidler, M.} \emph{et~al.}
\newblock \bibinfo{title}{{Fermi polaron-polaritons in charge-tunable
  atomically thin semiconductors}}.
\newblock \emph{\bibinfo{journal}{Nat. Phys.}} \textbf{\bibinfo{volume}{13}},
  \bibinfo{pages}{255--261} (\bibinfo{year}{2017}).
  
\bibitem{Efimkin2017}
\bibinfo{author}{Efimkin, D.~K.} \& \bibinfo{author}{MacDonald, A.~H.}
\newblock \bibinfo{title}{{Many-body theory of trion absorption features in
  two-dimensional semiconductors}}.
\newblock \emph{\bibinfo{journal}{Phys. Rev. B}} \textbf{\bibinfo{volume}{95}},
  \bibinfo{pages}{035417} (\bibinfo{year}{2017}).

\bibitem{Seyler2019}
\bibinfo{author}{Seyler, K.~L.} \emph{et~al.}
\newblock \bibinfo{title}{{Signatures of moir{\'{e}}-trapped valley excitons in
  MoSe2/WSe2 heterobilayers}}.
\newblock \emph{\bibinfo{journal}{Nature}} \textbf{\bibinfo{volume}{567}},
  \bibinfo{pages}{66--70} (\bibinfo{year}{2019}).

\bibitem{Tran2019}
\bibinfo{author}{Tran, K.} \emph{et~al.}
\newblock \bibinfo{title}{{Evidence for moir{\'{e}} excitons in van der Waals
  heterostructures}}.
\newblock \emph{\bibinfo{journal}{Nature}} \textbf{\bibinfo{volume}{567}},
  \bibinfo{pages}{71--75} (\bibinfo{year}{2019}).

\bibitem{Alexeev2019}
\bibinfo{author}{Alexeev, E.~M.} \emph{et~al.}
\newblock \bibinfo{title}{{Resonantly hybridized excitons in moir{\'{e}}
  superlattices in van der Waals heterostructures}}.
\newblock \emph{\bibinfo{journal}{Nature}} \textbf{\bibinfo{volume}{567}},
  \bibinfo{pages}{81--86} (\bibinfo{year}{2019}).

\bibitem{Jin2019}
\bibinfo{author}{Jin, C.} \emph{et~al.}
\newblock \bibinfo{title}{{Observation of moir{\'{e}} excitons in WSe2/WS2
  heterostructure superlattices}}.
\newblock \emph{\bibinfo{journal}{Nature}} \textbf{\bibinfo{volume}{567}},
  \bibinfo{pages}{76--80} (\bibinfo{year}{2019}).

\bibitem{Fogler2014}
\bibinfo{author}{Fogler, M.~M.}, \bibinfo{author}{Butov, L.~V.} \&
  \bibinfo{author}{Novoselov, K.~S.}
\newblock \bibinfo{title}{{High-temperature superfluidity with indirect
  excitons in van der Waals heterostructures}}.
\newblock \emph{\bibinfo{journal}{Nat. Commun.}} \textbf{\bibinfo{volume}{5}}
  (\bibinfo{year}{2014}).

\bibitem{Wang2019}
\bibinfo{author}{Wang, Z.} \emph{et~al.}
\newblock \bibinfo{title}{{Evidence of high-temperature exciton condensation in
  two-dimensional atomic double layers}}.
\newblock \emph{\bibinfo{journal}{Nature}} \textbf{\bibinfo{volume}{574}},
  \bibinfo{pages}{76--80} (\bibinfo{year}{2019}).

\bibitem{Deilmann2018}
\bibinfo{author}{Deilmann, T.} \& \bibinfo{author}{Thygesen, K.~S.}
\newblock \bibinfo{title}{{Interlayer Excitons with Large Optical Amplitudes in
  Layered van der Waals Materials}}.
\newblock \emph{\bibinfo{journal}{Nano Lett.}} \textbf{\bibinfo{volume}{18}},
  \bibinfo{pages}{2984--2989} (\bibinfo{year}{2018}).

\bibitem{Chaves2018}
\bibinfo{author}{Chaves, A.}, \bibinfo{author}{Azadani, J.~G.},
  \bibinfo{author}{{\"{O}}z{\c{c}}elik, V.~O.}, \bibinfo{author}{Grassi, R.} \&
  \bibinfo{author}{Low, T.}
\newblock \bibinfo{title}{{Electrical control of excitons in van der Waals
  heterostructures with type-II band alignment}}.
\newblock \emph{\bibinfo{journal}{Phys. Rev. B}} \textbf{\bibinfo{volume}{98}},
  \bibinfo{pages}{121302} (\bibinfo{year}{2018}).

\bibitem{Gerber2019}
\bibinfo{author}{Gerber, I.~C.} \emph{et~al.}
\newblock \bibinfo{title}{{Interlayer excitons in bilayer MoS2 with strong
  oscillator strength up to room temperature}}.
\newblock \emph{\bibinfo{journal}{Phys. Rev. B}} \textbf{\bibinfo{volume}{99}},
  \bibinfo{pages}{035443} (\bibinfo{year}{2019}).

\bibitem{Zheng1997}
\bibinfo{author}{Zheng, L.}, \bibinfo{author}{Ortalano, M.~W.} \&
  \bibinfo{author}{{Das Sarma}, S.}
\newblock \bibinfo{title}{{Exchange instabilities in semiconductor
  double-quantum-well systems}}.
\newblock \emph{\bibinfo{journal}{Phys. Rev. B}} \textbf{\bibinfo{volume}{55}},
  \bibinfo{pages}{4506--4515} (\bibinfo{year}{1997}).

\bibitem{Ezawa2000}
\bibinfo{author}{Ezawa, Z.~F.}
\newblock \emph{\bibinfo{title}{{Quantum Hall Effects: Field Theoretical
  Approach and Related Topics}}} (\bibinfo{publisher}{World Scientific},
  \bibinfo{year}{2000}).

\bibitem{Kim2016}
\bibinfo{author}{Kim, K.} \emph{et~al.}
\newblock \bibinfo{title}{{van der Waals Heterostructures with High Accuracy
  Rotational Alignment}}.
\newblock \emph{\bibinfo{journal}{Nano Lett.}} \textbf{\bibinfo{volume}{16}},
  \bibinfo{pages}{1989--1995} (\bibinfo{year}{2016}).

\bibitem{Zhang2014}
\bibinfo{author}{Zhang, Y.} \emph{et~al.}
\newblock \bibinfo{title}{{Direct observation of the transition from indirect
  to direct bandgap in atomically thin epitaxial MoSe2}}.
\newblock \emph{\bibinfo{journal}{Nat. Nanotechnol.}}
  \textbf{\bibinfo{volume}{9}}, \bibinfo{pages}{111--115}
  (\bibinfo{year}{2014}).

\bibitem{He2013}
\bibinfo{author}{He, K.}, \bibinfo{author}{Poole, C.}, \bibinfo{author}{Mak,
  K.~F.} \& \bibinfo{author}{Shan, J.}
\newblock \bibinfo{title}{{Experimental Demonstration of Continuous Electronic
  Structure Tuning via Strain in Atomically Thin MoS2}}.
\newblock \emph{\bibinfo{journal}{Nano Lett.}} \textbf{\bibinfo{volume}{13}},
  \bibinfo{pages}{2931--2936} (\bibinfo{year}{2013}).

\bibitem{Conley2013}
\bibinfo{author}{Conley, H.~J.} \emph{et~al.}
\newblock \bibinfo{title}{{Bandgap Engineering of Strained Monolayer and
  Bilayer MoS2}}.
\newblock \emph{\bibinfo{journal}{Nano Lett.}} \textbf{\bibinfo{volume}{13}},
  \bibinfo{pages}{3626--3630} (\bibinfo{year}{2013}).

\bibitem{Ozcelik2016}
\bibinfo{author}{{\"{O}}z{\c{c}}elik, V.~O.}, \bibinfo{author}{Azadani, J.~G.},
  \bibinfo{author}{Yang, C.}, \bibinfo{author}{Koester, S.~J.} \&
  \bibinfo{author}{Low, T.}
\newblock \bibinfo{title}{{Band alignment of two-dimensional semiconductors for
  designing heterostructures with momentum space matching}}.
\newblock \emph{\bibinfo{journal}{Phys. Rev. B}} \textbf{\bibinfo{volume}{94}},
  \bibinfo{pages}{035125} (\bibinfo{year}{2016}).

\bibitem{Xu2014}
\bibinfo{author}{Xu, X.}, \bibinfo{author}{Yao, W.}, \bibinfo{author}{Xiao, D.}
  \& \bibinfo{author}{Heinz, T.~F.}
\newblock \bibinfo{title}{{Spin and pseudospins in layered transition metal
  dichalcogenides}}.
\newblock \emph{\bibinfo{journal}{Nat. Phys.}} \textbf{\bibinfo{volume}{10}},
  \bibinfo{pages}{343--350} (\bibinfo{year}{2014}).

\bibitem{Back2017}
\bibinfo{author}{Back, P.} \emph{et~al.}
\newblock \bibinfo{title}{{Giant Paramagnetism-Induced Valley Polarization of
  Electrons in Charge-Tunable Monolayer MoSe2}}.
\newblock \emph{\bibinfo{journal}{Phys. Rev. Lett.}}
  \textbf{\bibinfo{volume}{118}}, \bibinfo{pages}{237404}
  (\bibinfo{year}{2017}).

\bibitem{Smolenski2019}
\bibinfo{author}{Smole{\'{n}}ski, T.} \emph{et~al.}
\newblock \bibinfo{title}{{Interaction-Induced Shubnikov–de Haas Oscillations
  in Optical Conductivity of Monolayer MoSe2}}.
\newblock \emph{\bibinfo{journal}{Phys. Rev. Lett.}}
  \textbf{\bibinfo{volume}{123}}, \bibinfo{pages}{097403}
  (\bibinfo{year}{2019}).

\bibitem{Hanson2007}
\bibinfo{author}{Hanson, R.}, \bibinfo{author}{Kouwenhoven, L.~P.},
  \bibinfo{author}{Petta, J.~R.}, \bibinfo{author}{Tarucha, S.} \&
  \bibinfo{author}{Vandersypen, L. M.~K.}
\newblock \bibinfo{title}{{Spins in few-electron quantum dots}}.
\newblock \emph{\bibinfo{journal}{Rev. Mod. Phys.}}
  \textbf{\bibinfo{volume}{79}}, \bibinfo{pages}{1217--1265}
  (\bibinfo{year}{2007}).

\bibitem{Eisenstein1994}
\bibinfo{author}{Eisenstein, J.~P.}, \bibinfo{author}{Pfeiffer, L.~N.} \&
  \bibinfo{author}{West, K.~W.}
\newblock \bibinfo{title}{{Compressibility of the two-dimensional electron gas:
  Measurements of the zero-field exchange energy and fractional quantum Hall
  gap}}.
\newblock \emph{\bibinfo{journal}{Phys. Rev. B}} \textbf{\bibinfo{volume}{50}},
  \bibinfo{pages}{1760--1778} (\bibinfo{year}{1994}).

\bibitem{Larentis2014}
\bibinfo{author}{Larentis, S.} \emph{et~al.}
\newblock \bibinfo{title}{{Band Offset and Negative Compressibility in
  Graphene-MoS2 Heterostructures}}.
\newblock \emph{\bibinfo{journal}{Nano Lett.}} \textbf{\bibinfo{volume}{14}},
  \bibinfo{pages}{2039--2045} (\bibinfo{year}{2014}).

\bibitem{Fallahazad2016}
\bibinfo{author}{Fallahazad, B.} \emph{et~al.}
\newblock \bibinfo{title}{{Shubnikov-de Haas Oscillations of High-Mobility
  Holes in Monolayer and Bilayer WSe2 : Landau Level Degeneracy, Effective
  Mass, and Negative Compressibility}}.
\newblock \emph{\bibinfo{journal}{Phys. Rev. Lett.}}
  \textbf{\bibinfo{volume}{116}}, \bibinfo{pages}{086601}
  (\bibinfo{year}{2016}).

\bibitem{Hunt2017}
\bibinfo{author}{Hunt, B.~M.} \emph{et~al.}
\newblock \bibinfo{title}{{Direct measurement of discrete valley and orbital
  quantum numbers in bilayer graphene}}.
\newblock \emph{\bibinfo{journal}{Nat. Commun.}} \textbf{\bibinfo{volume}{8}},
  \bibinfo{pages}{948} (\bibinfo{year}{2017}).

\bibitem{Imada1998}
\bibinfo{author}{Imada, M.}, \bibinfo{author}{Fujimori, A.} \&
  \bibinfo{author}{Tokura, Y.}
\newblock \bibinfo{title}{{Metal-insulator transitions}}.
\newblock \emph{\bibinfo{journal}{Rev. Mod. Phys.}}
  \textbf{\bibinfo{volume}{70}}, \bibinfo{pages}{1039--1263}
  (\bibinfo{year}{1998}).

\bibitem{Camjayi2008}
\bibinfo{author}{Camjayi, A.}, \bibinfo{author}{Haule, K.},
  \bibinfo{author}{Dobrosavljevi{\'{c}}, V.} \& \bibinfo{author}{Kotliar, G.}
\newblock \bibinfo{title}{{Coulomb correlations and the Wigner–Mott
  transition}}.
\newblock \emph{\bibinfo{journal}{Nat. Phys.}} \textbf{\bibinfo{volume}{4}},
  \bibinfo{pages}{932--935} (\bibinfo{year}{2008}).

\bibitem{Zarenia2017}
\bibinfo{author}{Zarenia, M.}, \bibinfo{author}{Neilson, D.} \&
  \bibinfo{author}{Peeters, F.~M.}
\newblock \bibinfo{title}{{Inhomogeneous phases in coupled electron-hole
  bilayer graphene sheets: Charge Density Waves and Coupled Wigner Crystals}}.
\newblock \emph{\bibinfo{journal}{Sci. Rep.}} \textbf{\bibinfo{volume}{7}},
  \bibinfo{pages}{11510} (\bibinfo{year}{2017}).

\bibitem{Ludwig2011}
\bibinfo{author}{Ludwig, D.}, \bibinfo{author}{Floerchinger, S.},
  \bibinfo{author}{Moroz, S.} \& \bibinfo{author}{Wetterich, C.}
\newblock \bibinfo{title}{{Quantum phase transition in Bose-Fermi mixtures}}.
\newblock \emph{\bibinfo{journal}{Phys. Rev. A}} \textbf{\bibinfo{volume}{84}},
  \bibinfo{pages}{033629} (\bibinfo{year}{2011}).

\bibitem{Little1964}
\bibinfo{author}{Little, W.~A.}
\newblock \bibinfo{title}{{Possibility of Synthesizing an Organic
  Superconductor}}.
\newblock \emph{\bibinfo{journal}{Phys. Rev.}} \textbf{\bibinfo{volume}{134}},
  \bibinfo{pages}{A1416--A1424} (\bibinfo{year}{1964}).

\bibitem{Ginzburg1970}
\bibinfo{author}{Ginzburg, V.~L.}
\newblock \bibinfo{title}{{THE PROBLEM OF HIGH-TEMPERATURE SUPERCONDUCTIVITY.
  II}}.
\newblock \emph{\bibinfo{journal}{Sov. Phys. Uspekhi}}
  \textbf{\bibinfo{volume}{13}}, \bibinfo{pages}{335--352}
  (\bibinfo{year}{1970}).

\bibitem{Laussy2010}
\bibinfo{author}{Laussy, F.~P.}, \bibinfo{author}{Kavokin, A.~V.} \&
  \bibinfo{author}{Shelykh, I.~A.}
\newblock \bibinfo{title}{{Exciton-Polariton Mediated Superconductivity}}.
\newblock \emph{\bibinfo{journal}{Phys. Rev. Lett.}}
  \textbf{\bibinfo{volume}{104}}, \bibinfo{pages}{106402}
  (\bibinfo{year}{2010}).

\bibitem{Cotlet2016}
\bibinfo{author}{Cotlet, O.}, \bibinfo{author}{Zeytinoglu, S.},
  \bibinfo{author}{Sigrist, M.}, \bibinfo{author}{Demler, E.} \&
  \bibinfo{author}{Imamoglu, A.}
\newblock \bibinfo{title}{{Superconductivity and other collective phenomena in
  a hybrid Bose-Fermi mixture formed by a polariton condensate and an electron
  system in two dimensions}}.
\newblock \emph{\bibinfo{journal}{Phys. Rev. B}} \textbf{\bibinfo{volume}{93}},
  \bibinfo{pages}{054510} (\bibinfo{year}{2016}).

\bibitem{Wang2013}
\bibinfo{author}{Wang, L.} \emph{et~al.}
\newblock \bibinfo{title}{{One-Dimensional Electrical Contact to a
  Two-Dimensional Material}}.
\newblock \emph{\bibinfo{journal}{Science}} \textbf{\bibinfo{volume}{342}},
  \bibinfo{pages}{614--617} (\bibinfo{year}{2013}).

\end{thebibliography}


\begin{thebibliography}{1}
\expandafter\ifx\csname url\endcsname\relax
  \def\url#1{\texttt{#1}}\fi
\expandafter\ifx\csname urlprefix\endcsname\relax\def\urlprefix{URL }\fi
\providecommand{\bibinfo}[2]{#2}
\providecommand{\eprint}[2][]{\url{#2}}

\bibitem{Sidler2017}
\bibinfo{author}{Sidler, M.} \emph{et~al.}
\newblock \bibinfo{title}{{Fermi polaron-polaritons in charge-tunable
  atomically thin semiconductors}}.
\newblock \emph{\bibinfo{journal}{Nat. Phys.}} \textbf{\bibinfo{volume}{13}},
  \bibinfo{pages}{255--261} (\bibinfo{year}{2017}).

\bibitem{Smolenski2019}
\bibinfo{author}{Smole{\'{n}}ski, T.} \emph{et~al.}
\newblock \bibinfo{title}{{Interaction-Induced Shubnikov–de Haas Oscillations
  in Optical Conductivity of Monolayer MoSe2}}.
\newblock \emph{\bibinfo{journal}{Phys. Rev. Lett.}}
  \textbf{\bibinfo{volume}{123}}, \bibinfo{pages}{097403}
  (\bibinfo{year}{2019}).

\bibitem{Back2017}
\bibinfo{author}{Back, P.} \emph{et~al.}
\newblock \bibinfo{title}{{Giant Paramagnetism-Induced Valley Polarization of
  Electrons in Charge-Tunable Monolayer MoSe2}}.
\newblock \emph{\bibinfo{journal}{Phys. Rev. Lett.}}
  \textbf{\bibinfo{volume}{118}}, \bibinfo{pages}{237404}
  (\bibinfo{year}{2017}).

\end{thebibliography}

\vspace{1 cm}

\section*{Acknowledgments}
We acknowledge discussions with E. Demler, R. Schmidt, T. Smolenski, A. Popert, and P. Kn\"{u}ppel.
This work was supported by the Swiss National Science Foundation (SNSF) under
Grant No. 200021-178909/1, the Japan Society for the Promotion of Science (JSPS) Postdoctoral Fellowship for Research Abroad.\\

\section*{Author Contributions} Y.S. and I.S. carried out the measurements. Y.S. designed and fabricated the sample.
M.K. helped to prepare the experimental setup.
K.W. and T.T. grew hBN crystal.
Y.S., I.S. and A.I. wrote the manuscript.
A.I. supervised the project.\\

\section*{Competing interests} The authors declare no competing financial interests.

\section*{Additional information}
{\bf Supplementary information} for this paper is available online.

{\bf Correspondence and requests for materials} should be addressed to A.I. (imamoglu@phys.ethz.ch).

\end{document}

\baselineskip24pt

\maketitle

\section{Differential reflectance and Photoluminescence of monolayer {$\bf MoSe_2$}}
\begin{figure*}[h!]
\centering
\includegraphics[width=1.0\textwidth]{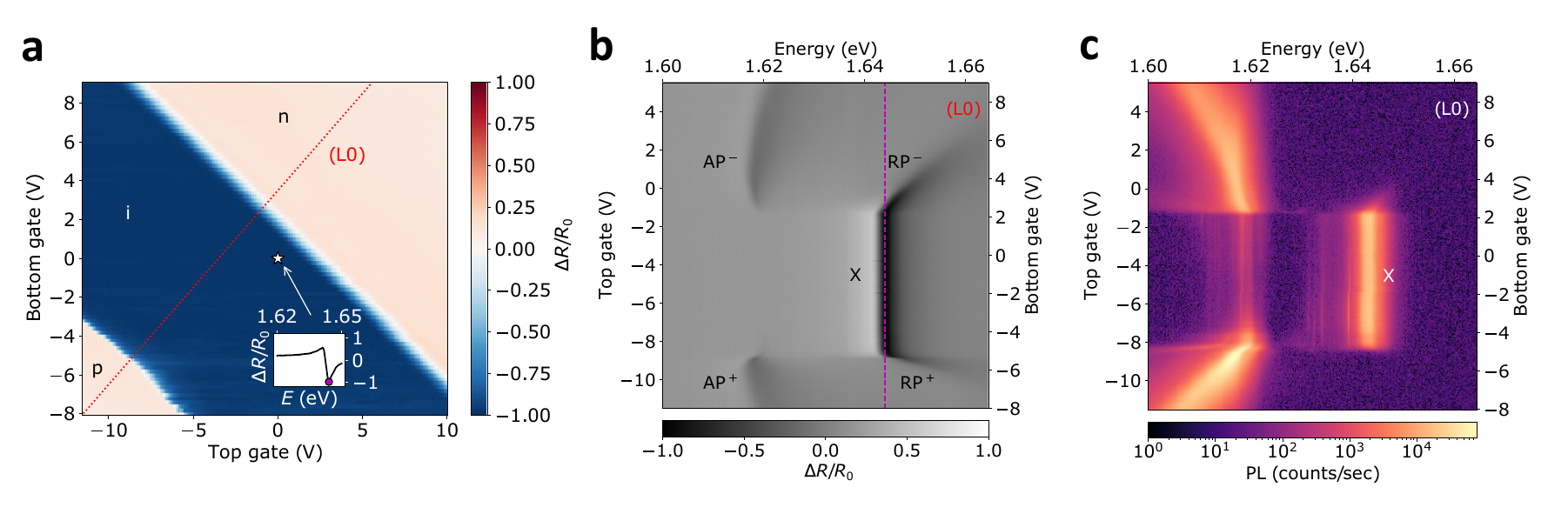}
\caption{{\bf Gate dependence of differential reflectance spectrum and PL of $\bf MoSe_2$.}
{\bf a} Two gates dependence map of differential reflectance
around exciton resosnace ($E = 1.644 \rm eV$).
The charge configuration is indicated by p, i, n
which correspond to hole doped, neutral, electron doped, respectively.
The inset shows the differential reflectance spectrum at $(V_{\rm tg}, V_{\rm bg}) = (0{\rm V}, 0{\rm V})$ 
(indicated with the white star in the map).
The magenta dot in the inset indicate the point where $E = 1.644 \rm eV$.
{\bf b} Gate dependence of differential reflectance along the red dashed line L0 shown in {\bf a}.
Magenta dashed line indicates the exciton resonance energy ($E = 1.644 \rm eV$).
{\bf c} Gate dependence of PL along the red dashed line L0 shown in {\bf a}.
}
\label{figS1}
\end{figure*}

Here we show gate dependence of differential reflectance and photoluminescence (PL) of monolayer \MoSe.
Fig. \ref{figS1}a shows the gate dependence of the differential reflectance at $E = 1.644 {\rm eV}$.
This energy corresponds to the minimum of the exciton reflection spectrum, as shown in the inset of Fig. \ref{figS1}a. In contrast to the case of $\rm MoSe_2/hBN/MoSe_2$ (Figs. 2a and 2b in the main text),
the dependences on top gate and bottom gate are similar.
The blue area where we have absorption from the exciton resonance is where \MoSe\ is neutral.
In Fig. \ref{figS1}b,
we plot the gate dependence of the differential reflectance spectrum along L0 in Fig. \ref{figS1}a.
When \MoSe\ is doped, the exciton resonance around $E = 1.644 {\rm eV}$ becomes the repulsive polaron and blue shifts. This is accompanied by the emergence of the attractive polaron\cite{Sidler2017} resonance around $E = 1.619 {\rm eV}$. Due to this blue shift of the exciton resonance
together with the dispersive shape of the exciton resonance as shown in the inset of Fig. \ref{figS1}a,
we can detect the changes in the charge configuration as we show in Fig. \ref{figS1}a.
In Fig. \ref{figS1}c, we plot gate dependence of the PL spectrum along L0 in Fig. \ref{figS1}a.
We observe that exciton PL around $E = 1.644 {\rm eV}$ is prominent in the neutral regime,
while attractive polaron PL around $E = 1.619 {\rm eV}$ dominates in the electron and hole doped regimes.

\newpage

\section{Differential reflectance and Photoluminescence of $\bf MoSe_2/hBN/MoSe_2$ at a different spot}
In the main text, we show the spectroscopy data of $\rm MoSe_2/hBN/MoSe_2$ at a specific spot.
Here we show  charge configuration detection and inter- and intra-layer exciton coupling at a different spot in the $\rm MoSe_2/hBN/MoSe_2$ section of the sample.

\begin{figure*}[h!]
	\centering
	\includegraphics[width=1.0\textwidth]{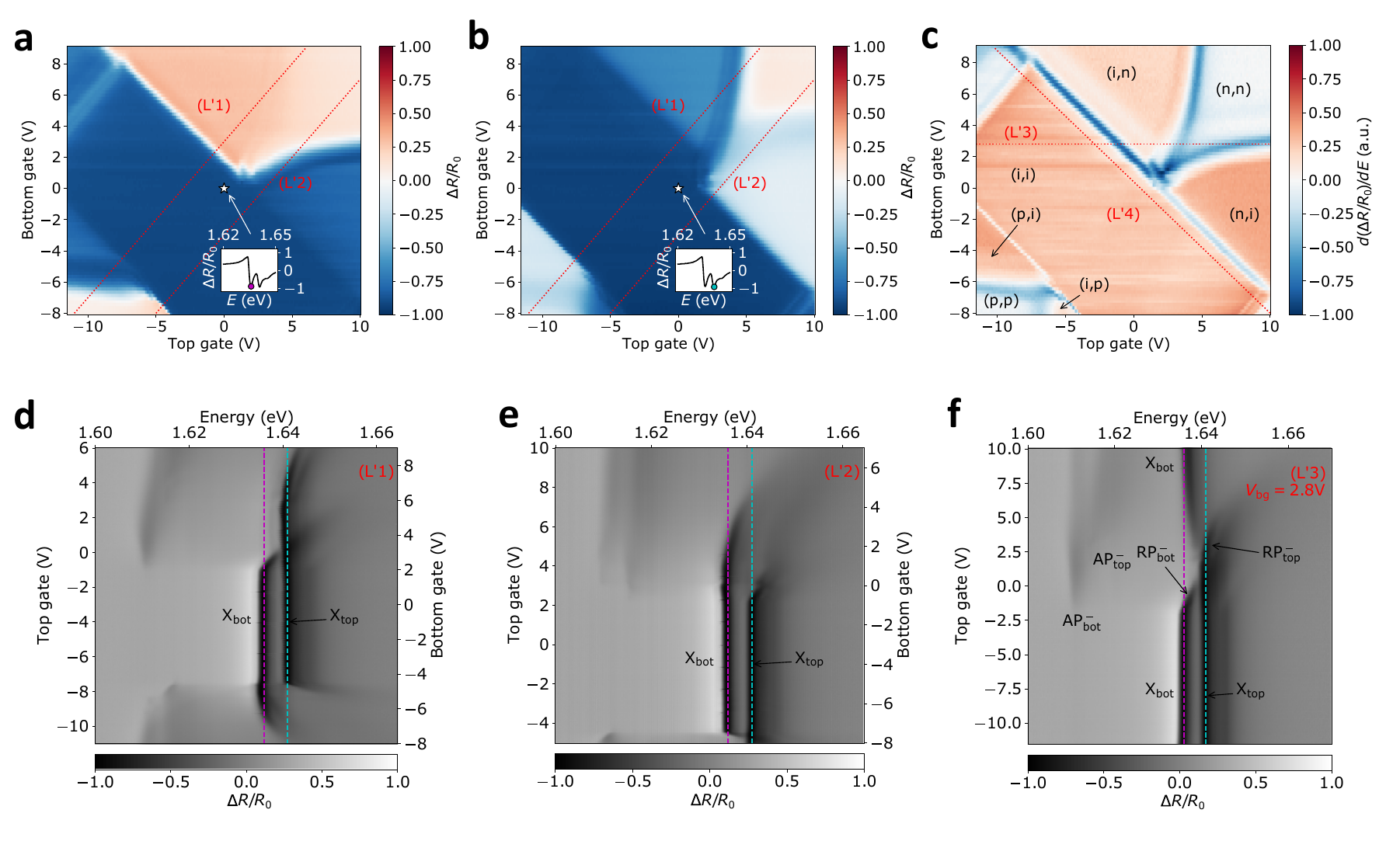}
	\caption{{\bf Gate dependence of differential reflectance spectrum of $\bf MoSe_2/hBN/MoSe_2$ at a different spot.}
		{\bf a, b} Two gates dependence maps of differential reflectance
		around bottom ({\bf a}) and top ({\bf b}) intra-layer exciton resosnaces
		($E = 1.636 \rm eV$ and $E = 1.641 \rm eV$, respectively).
		The insets of {\bf a} and {\bf b} show the differential reflectance spectrum
		at $(V_{\rm tg}, V_{\rm bg}) = (0{\rm V}, 0{\rm V})$
		(indicated with the white stars in the maps).
		The magenta and cyan dots in the insets indicate the points
		where $E = 1.636 \rm eV$ and $E = 1.641 \rm eV$, respectively.
		{\bf c} Charge configuration diagram obtained by derivative of the differential reflectance spectrum
		with respect to energy (sum of the derivatives at $E = 1.636 \rm eV$ and $E = 1.641 \rm eV$).
		The charge configuration for each layer is indicated by p, i, n
		which correspond to hole doped, neutral, electron doped, respectively,
		and shown in the order of (top, bottom).
		{\bf d} - {\bf f} Gate dependence of differential reflectance along the red dashed lines L'1 ({\bf d}), L'2 ({\bf e}), and L'3 ({\bf f})
		shown in {\bf a}, {\bf b} and {\bf c}.
		Magenta and cyan dashed lines indicate the bottom ($E = 1.636 \rm eV$) and top ($E = 1.641 \rm eV$)
		exciton resonance energies, respectively.
	}
	\label{figS2}
\end{figure*}

Fig. \ref{figS2}a and \ref{figS2}b show gate dependence the differential reflectance
at $E = 1.636 {\rm eV}$ and $E = 1.641 {\rm eV}$, respectively.
These energies correspond to minima of the exciton resonances in the charge neutral regime 
(see the insets of Fig. \ref{figS2}a and \ref{figS2}b).
As we discussed in the main text and also in section S1,
carrier doping induces the blue shift of the exciton resonances
due to the formation of repulsive polaron,
which results in the abrupt enhancement of the differential reflectance
in Fig. \ref{figS2}a and \ref{figS2}b.
From the gate dependence,
we assign the exciton resonances at $E = 1.636 {\rm eV}$ and $E = 1.641 {\rm eV}$
to the bottom and the top intra-layer excitons, respectively.
Note that the energy order of top and bottom intra-layer excitons is opposite to
the result shown in the main text.
This is most likely caused by the inhomogeneity of the strain across the sample.
To enhance the border of the transition and visualize the charge configuration of both layers at the same time,
we calculated the derivative of the differential reflectance signal  
with respect to energy at $E = 1.636 \rm eV$ and $E = 1.641 \rm eV$,
and added them together (Fig. \ref{figS2}c).
In Fig. \ref{figS2}d, e, and f, we show the gate dependence of the differential reflectance spectrum
along L'1, L'2 and L'3 (see Figs. \ref{figS2}a, \ref{figS2}b and \ref{figS2}c), respectively.
The results are similar to the results we show in the main text including the signature of negative compressibility. The only qualitative difference with respect to the data in the main text is the opposite energy order
of the top and the bottom intra-layer exciton resonances.

\begin{figure*}[h!]
\centering
\includegraphics[width=1.0\textwidth]{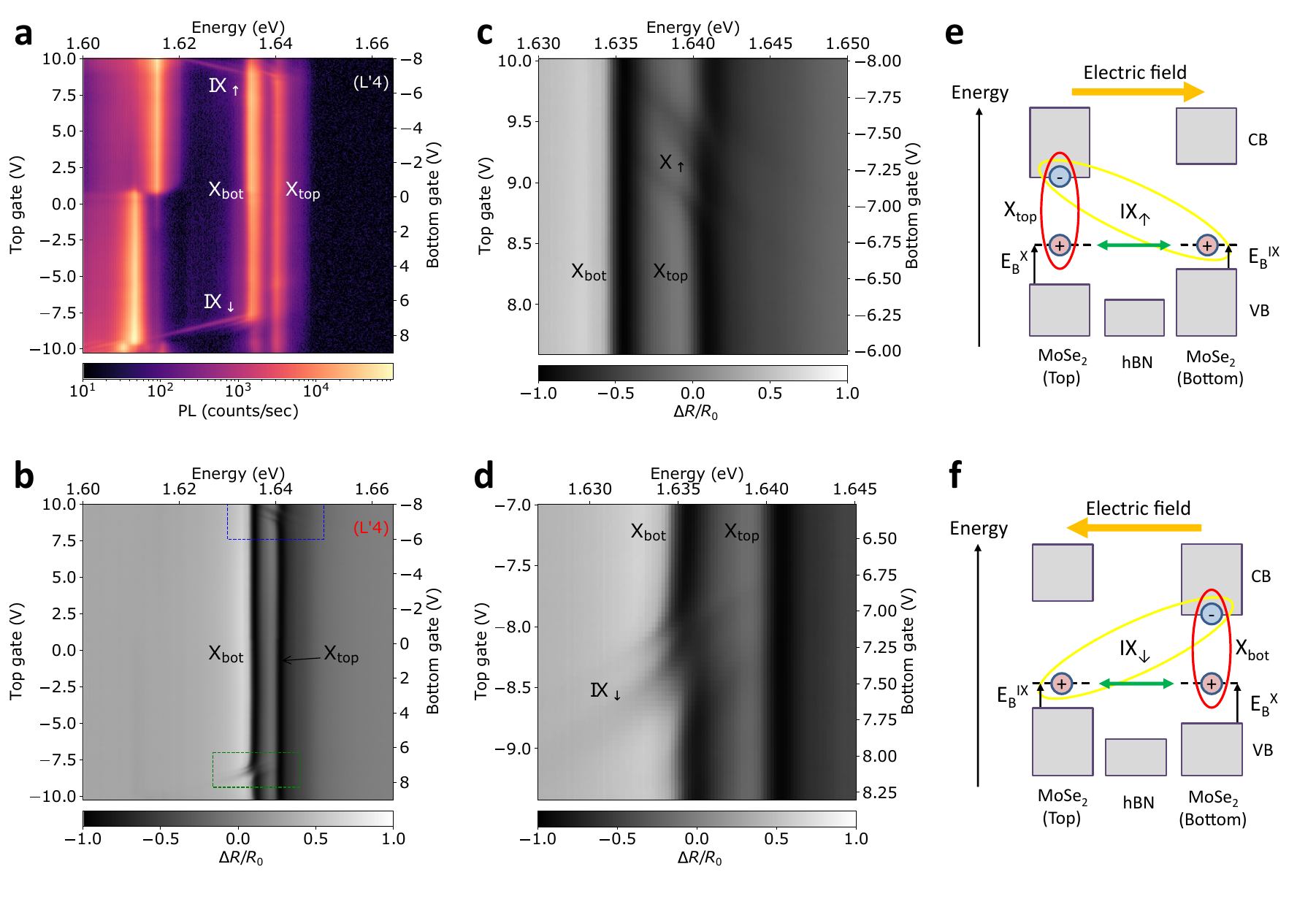}
\caption{{\bf Electric field dependence of PL and differential reflectance at charge neutrality.}
Gate dependence of PL ({\bf a}) and differential reflectance ({\bf b}) of $\rm MoSe_2/hBN/MoSe_2$ 
at the different spot along the dashed line L'4 shown in Fig. \ref{figS2}c.
Top and bottom gate voltages are scanned together to tune the electric field but to fix the chemical potential.
The intensity of the PL ({\bf a}) is shown in log scale.
{\bf c, d} Magnified plots of {\bf b}.
The corresponding area of {\bf c} and {\bf d} is indicated by the blue and green dashed rectangulars
in {\bf b}, respectively.
{\bf e, f} Schematic image of the energy band and the exciton energy alignment under electric fields
when the resonant conditions are achieved.
{\bf e} and {\bf f} correspond to the situation of {\bf c} and {\bf d}, respectively.
}
\label{figS3}
\end{figure*}

We also show the electric field dependence of PL (Fig. \ref{figS3}a) 
and differential reflectance (Fig. \ref{figS3}b)
in the charge neutrality regime at this spot.
For both measurements, the top and bottom gates are scanned along L'4 indicated in Fig. \ref{figS2}c
to control the electric field $E_z$ while keeping the system at charge neutrality.
In Fig. \ref{figS3}a, there are inter-layer exciton PL lines
which show Stark shift in the high $E_z$ regime
(top and botton side of the figure).
In Fig. \ref{figS3}b, we also observe differential reflection from the inter-layer excitons, exhibiting strong Stark shift, for high $E_z$. Fig. \ref{figS3}c and \ref{figS3}d show the magnified plots
of the differential reflection for high $|E_z|$ where we observe the inter-layer exciton reflection,
and we find avoided crossing structure in both plots. In Fig. \ref{figS3}c, we find the avoided crossings
between the top intra-layer exciton ($E\sim 1.641 \rm eV$) and the inter-layer excitons 
 with an electron in the top and a hole in the bottom layer,
evidencing the coupling of these states via coherent hole tunneling (Fig. \ref{figS3}e).
In Fig. \ref{figS3}d, we find the avoided crossings
between the bottom intra-layer exciton ($E\sim 1.636 \rm eV$) and the inter-layer excitons 
with the opposite dipole, once again confirming that the coherent coupling of these states via hole tunneling (Fig. \ref{figS3}f).

All of the results are fully consistent with the results we show in the main text,
even though the energy order of top and bottom intra-layer excitons is opposite.
Note that even the fine structure of the inter-layer exciton in Fig. \ref{figS3}c and Fig. \ref{figS3}d
is similar to the result which we show in the main text (Fig. 2c and 2d). This observation indicates that the origin of the fine structure is intrinsic to the system  and is consistent with \Moire\ potential picture. In particular, it shows that multiple inter-exciton lines do not stem from random spatial inhomogeneities in the sample. 

\newpage

\section{Detail of reflectance spectrum fitting and additional spectroscopy data in the low electron density regime}

We define the differential reflectance as $\Delta R/R_0 \equiv (R - R_0)/R_0$ where $R_0$ is the background reflectance, measured in a region of the heterostructure without \MoSe \space flakes, and $R$ is the experimentally measured reflectance signal. In this work we are interested in the reflectance from \MoSe/hBN/\MoSe \space layers. In order to account for reflections and losses from multiple layers of the heterostructure, we assumed a Lorentzian lineshape for the excitons $\chi(E)\propto \frac{\gamma}{E-E_0+i\gamma/2}$ and effectively describe the total reflectance by ${
\rm Im}[\chi(E)\exp(i\alpha)]$ \cite{Smolenski2019}. Here, $\alpha$ is a phase shift that depends in energy and gate voltage. The total reflectance signal is given by

\begin{equation}
\Delta R/R_0 = A\cos(\alpha)\frac{\gamma^2/2}{(E-E_0)^2+\gamma^2/4}-A\sin(\alpha)\frac{\gamma (E-E_0)}{(E-E_0)^2+\gamma^2/4} + C ,
\label{eqS1}
\end{equation}
where we introduced the parameter $C$ to capture the broad background signal.

\begin{figure*}[h!]
	\centering
	\includegraphics[width=0.65\textwidth]{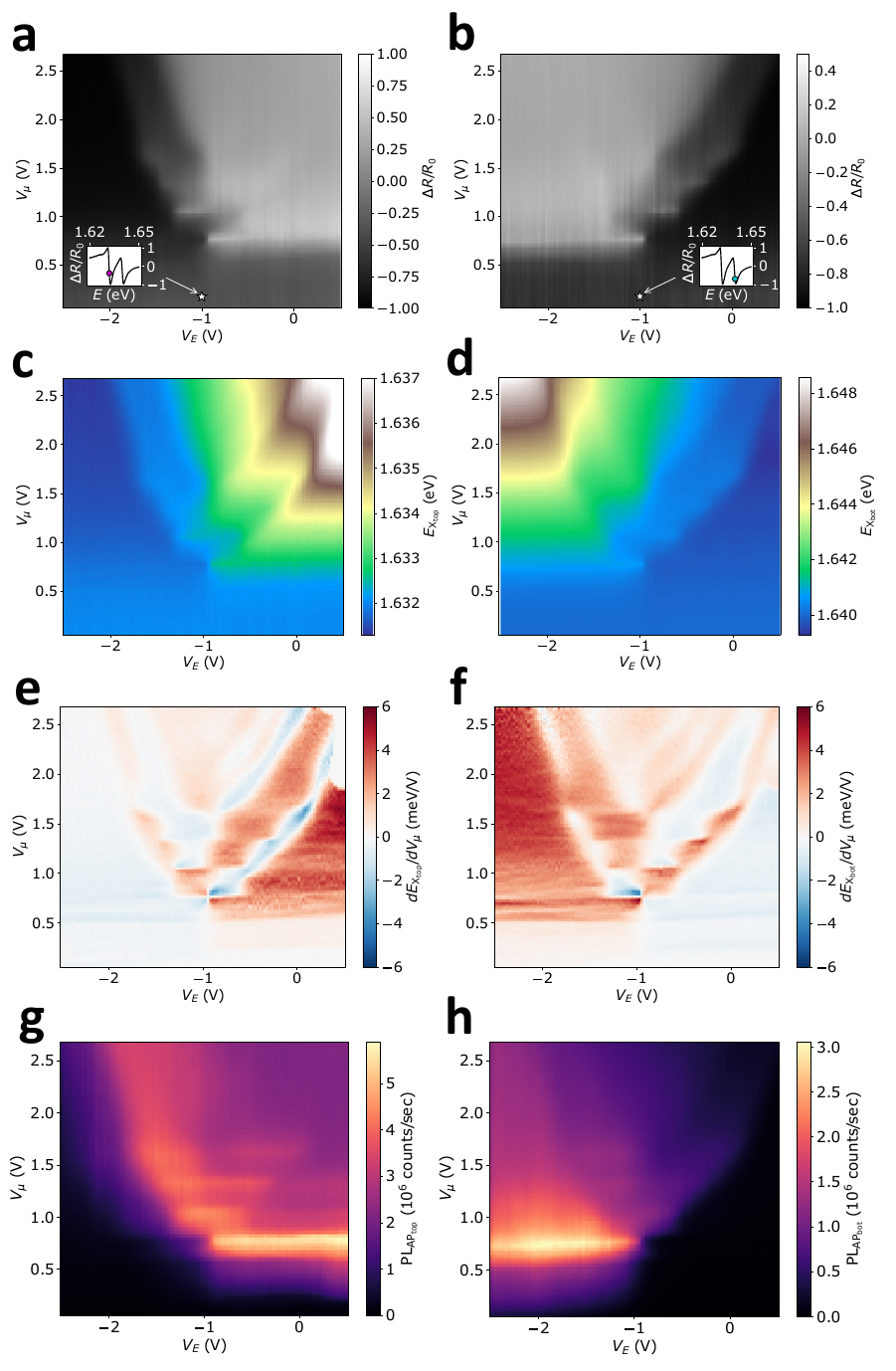}
	\caption{{\bf Spectroscopy data of low electron density regime.} 
		{\bf a, b} Gate dependence maps of differential reflectance
		around top ({\bf a}) and bottom ({\bf b}) intra-layer exciton resonances
		($E = 1.6320 \rm eV$ and $E = 1.6402 \rm eV$, respectively).  The insets of ({\bf a}) and ({\bf b}) show the differential reflectance spectrum
		at $(V_{E}, V_{\mu}) = (-1{\rm V}, 0.175{\rm V})$ (indicated with the white stars in the maps).
		{\bf c, d} Gate dependence maps of top ({\bf c}) and bottom ({\bf d}) intra-layer exciton resonance energy extracted from fitting of differential reflectance data.
		{\bf e, f} Gate dependence maps of top ({\bf e}) and bottom ({\bf f}) intra-layer exciton resonance energy differentiated by $V_{\mu}$.
		{\bf g, h} Gate dependence maps of top ({\bf g}) and bottom ({\bf h}) attractive polaron PL integrated over 1.6000 to 1.6083eV and 1.6125 to 1.6167eV, respectively. 
	}
	\label{figS4}
\end{figure*}

In Fig. \ref{figS4}a and \ref{figS4}b we show once again the gate voltage dependent maps of differential reflectance around top and bottom excitons used in Fig. 4a and 4b of the main text (where the gate voltages axes are $V_E = (7/15) V_{\rm tg} + (8/15) V_{\rm bg}$
and $V_{\mu} = 0.5 V_{\rm tg} - 0.5 V_{\rm bg}$). We used eq. \ref{eqS1} to fit the data and plotted the exciton resonance energies as function of gate voltages in Fig. \ref{figS4}c and \ref{figS4}d. The structures observed in Fig. \ref{figS4}a and \ref{figS4}b are well reproduced in Fig. \ref{figS4}c and \ref{figS4}d, respectively.
Figures \ref{figS4}e and \ref{figS4}f (and Fig. 4c and 4d in the main text) shows the derivative of the fitted exciton resonance from Fig. \ref{figS4}c and \ref{figS4}d with respect to $V_\mu$. The complementary checkerboard patterns are discussed in the main text, confirming the interpretation of layer by layer filling of electrons with changing chemical potential.
Periodic structures also appear in the integrated PL intensity of top and bottom layer attractive polaron (Fig. \ref{figS4}g and \ref{figS4}h, respectively), which is enhanced at the half integer filling of $\nu = $1/2, 1, 3/2 and 2. We indicate the charge configuration with ($\nu_{\rm top},\nu_{\rm bot}$), with $\nu_{\rm top} + \nu_{\rm bot} = \nu$.

\begin{figure*}[h!]
\centering
\includegraphics[width=1.0\textwidth]{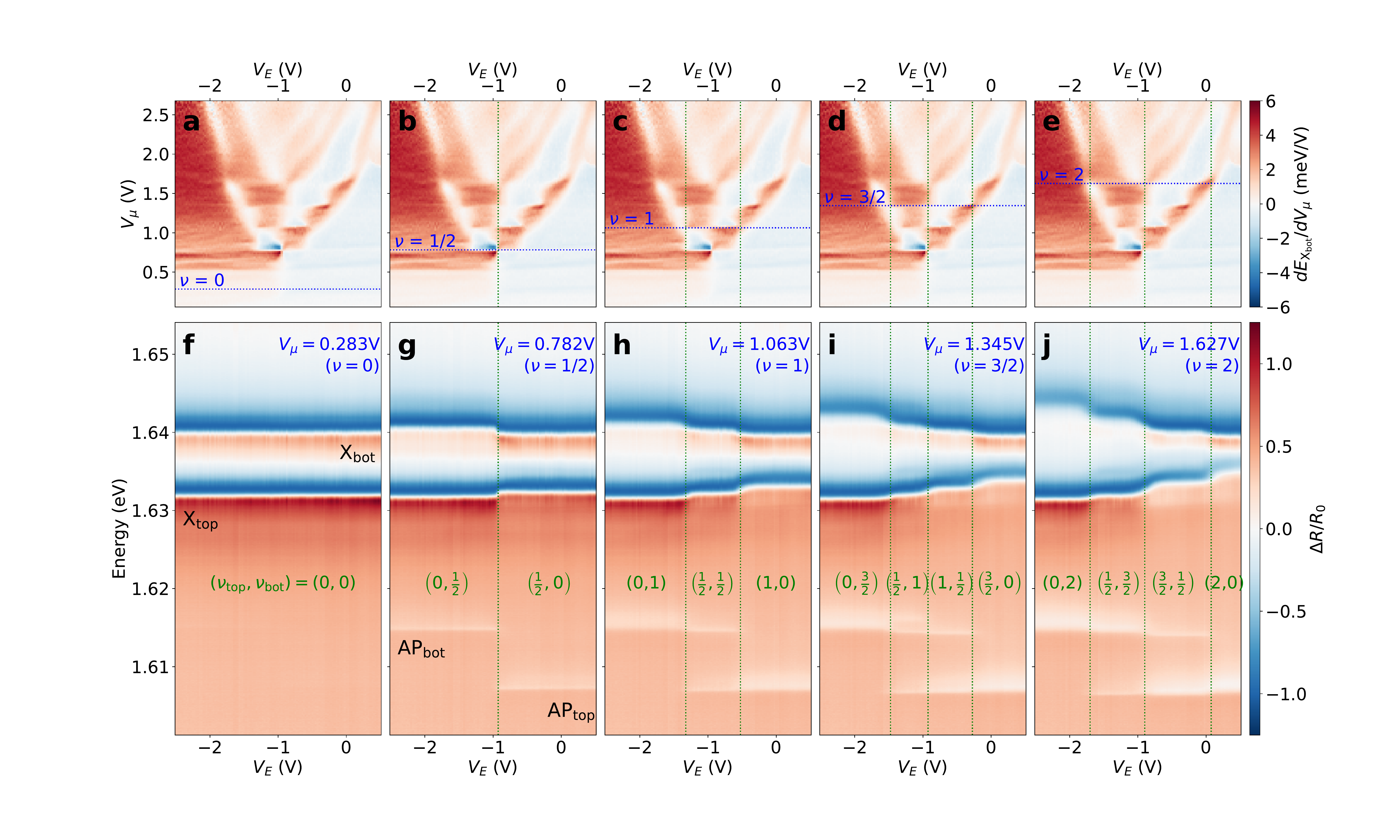}
\caption{{\bf Differential reflectance spectrum in low electron density regime.}
{\bf a} - {\bf e} Diagrams indicating electron filling of bottom layer (Fig. S4f).
{\bf f} - {\bf j} $V_E$ dependence of differential reflectance spectrum
for each fixed $V_{\mu}$ indicated by the blue broken lines in {\bf a} to {\bf e}.
Charging configuration of top and bottom layer is indicated by 
$(\nu_{\rm top}, \nu_{\rm bot})$.
Green broken lines indicate the transition point of charging configuration,
which are also indicated in {\bf a} to {\bf e}.
}
\label{figS5}
\end{figure*}

Fig. \ref{figS5}f - g shows additional filling factors, complementing Fig. 5 of the main text; here, we plot $V_E$ dependence of differential reflectance spectrum for fixed $V_{\mu}$ where $\nu = $0, 1/2, 1, 3/2, and 2 ($\nu = 1/2$ and $\nu = 1$ are presented in Fig 5. in the main text).
The $V_{\mu}$ (or $\nu$) values corresponding to Fig. \ref{figS5}f - g are indicated by a blue dotted lines
shown in Fig. \ref{figS5}a - e, showing $dE_{\rm X_{bot}}/dV_{\mu}$ (Fig. \ref{figS4}f).
For $\nu = 0$ (Fig. \ref{figS5}f), both layers are neutral and therefore $\rm X_{top}$ and $\rm X_{bot}$ resonances do not shift. In stark contrast to (1/2, 1/2) state at $\nu = 1$ filling, at $\nu = 2$ filling the (1, 1) state is missing, indicating that the state of integer filling in both layers is not sufficiently stabilized by the interactions.

\newpage

\section{Differential reflectance spectrum in the low electron density regime under a perpendicular magnetic field}
\begin{figure*}[h!]
\centering
\includegraphics[width=1.0\textwidth]{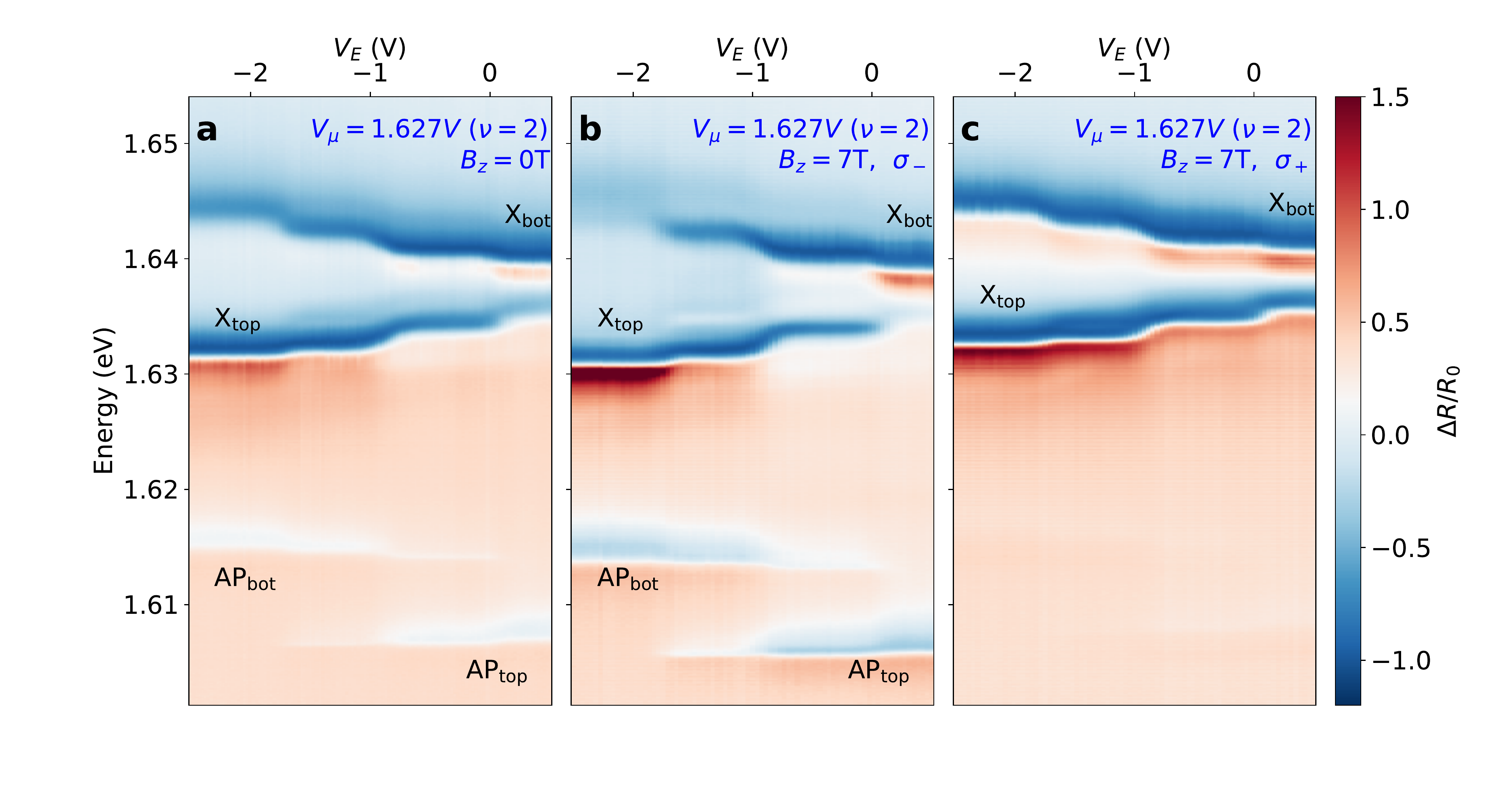}
\caption{{\bf Differential reflectance spectrum under perpendicular magnetic field.}
{\bf a} - {\bf c} $V_E$ dependence of differential reflectance spectrum
at $\nu = 2$ ($V_{\mu} = 1.627{\rm V})$.
{\bf a} is the reflectance spectrum at $B_z = 0{\rm T}$.
{\bf b} and {\bf c} are the reflectance spectrum at $B_z = 7{\rm T}$ in $\sigma_-$ and $\sigma_+$ circular polarization, respectively.
} 
\label{figS6}
\end{figure*}

Here we show $V_E$ dependence of
the differential reflectance spectrum in the low electron density regime
(at $\nu = 2$) obtained under perpendicular magnetic fleld ($B_z$).
Figs. \ref{figS6}b and \ref{figS6}c show the differential reflectance specturm obtained at $B_z = 7{\rm T}$ for $\sigma_{-}$ and $\sigma_{+}$ circular polarization, respectively.
In Fig. \ref{figS6}b, attractive polaron (AP) signal is clearly visible, but is missing in Fig. \ref{figS6}c.
$\sigma_{+}$ ($\sigma_{-}$) circularly polarized exciton-polaron is formed between K (-K) valley exciton and -K (K) valley Fermi sea electrons \cite{Back2017, Smolenski2019}.
The full $\sigma_{-}$ polarization of AP reflectance demonstrates that 
Fermi sea electrons are fully K-valley polarized at $B_z = 7{\rm T}$, $\nu = 2$.

Remarkably, even though Fermi sea electrons are fully valley polarized and the degeneracy of the \Moire\ subbands is reduced by a factor of 2, the structure of the spectrum remains essentially the same as the case of $B_z = 0{\rm T}$ (Fig. \ref{figS6}a). Our experiments therefore show that it is the number of electrons in \Moire\ unit cell that determines the structure, rather than the degeneracy of the quantum number. This observation is fully consistent with what we would expect from incompressible states induced by electron-electron interactions, where the degeneracy of \Moire\ subbands is lifted by strong interactions.

\newpage

\bibliographystyle{naturemag.bst}
\bibliography{References}

%\begin{thebibliography}{}
%\bibitem{Sidler2017}  Sidler, M. \textit{et al.} Fermi polaron-polaritons in charge-tunable atomically thin semiconductors. \textit{Nat. Phys.} \textbf{13}, 255--261 (2017).

%\end{thebibliography}

\clearpage